\documentclass[11pt,twocolappendix]{emulateapj}
\usepackage{amssymb,amsmath}
\usepackage{graphicx}
\usepackage{natbib}
\usepackage{epstopdf}
\usepackage{array}
\newcolumntype{$}{>{\global\let\currentrowstyle\relax}}
\newcolumntype{^}{>{\currentrowstyle}}

\def\etal{{\it \etal\ }} 

\def\eg{{\it e.g.\ }}

\def\p3m{P${}^3$M} 
\def\ap3m{AP${}^3$M} \def\-{{\em{---}}}

\def\gsim{\;\rlap{\lower 2.5pt
\hbox{$\sim$}}\raise 1.5pt\hbox{$>$}\;}
\def\lsim{\;\rlap{\lower 2.5pt
\hbox{$\sim$}}\raise 1.5pt\hbox{$<$}\;}
\def\etal{{\it et al\ }}
\def\eg{{\it e.g.\ }}
\newcommand{\be}{\begin{equation}}  \newcommand{\ba}{\begin{eqnarray}}
\newcommand{\ee}{\end{equation}}  \newcommand{\ea}{\end{eqnarray}}

 \newcommand{\bi}{\begin{itemize}}
\newcommand{\ei}{\end{itemize}}

\def\lesssim{\mathrel{\hbox{\rlap{\hbox{\lower4pt\hbox{$\sim$}}}\hbox{$<$}}}}
\def\gtrsim{\mathrel{\hbox{\rlap{\hbox{\lower4pt\hbox{$\sim$}}}\hbox{$>$}}}}

\begin{document}

\title{The Launching of Cold Clouds by Galaxy Outflows I: \\ Hydrodynamic Interactions with Radiative Cooling}
\author{Evan Scannapieco\altaffilmark{1} \& Marcus Br\"uggen\altaffilmark{2}}
\altaffiltext{1}{School of Earth and Space Exploration,  Arizona State University, P.O.  Box 871404, Tempe, AZ, 85287-1404}
\altaffiltext{2}{Universitat Hamburg, Hamburger Sternwarte, Gojenbergsweg 112, 21029, Hamburg, Germany}

\begin{abstract}

To better understand the nature of the multiphase material found in outflowing galaxies, we study the evolution of cold clouds embedded in flows of hot and fast material. Using a suite of adaptive-mesh refinement simulations that include radiative cooling, we investigate both cloud mass loss and cloud acceleration under the full range of conditions observed in galaxy outflows. The simulations are designed to track the cloud center of mass, enabling us to study the cloud evolution at long disruption times.  For supersonic flows, a Mach cone forms around the cloud, which damps the Kelvin-Helmholtz instability but also establishes a streamwise pressure gradient that stretches the cloud apart. If time is expressed in units of the cloud crushing time, both the cloud lifetime and the cloud acceleration rate are independent of cloud radius, and we find simple scalings for these quantities as a function of the Mach number of the external medium. A resolution study suggests that our simulations have sufficient resolution to accurately describe the evolution of cold clouds in the absence of thermal conduction and magnetic fields, physical processes whose roles will be studied in forthcoming papers.

\end{abstract}

\section{Introduction}

Galaxy outflows occur in rapidly star forming galaxies of all masses and at all redshifts,  and they  play a central role in the history of galaxy formation  (\eg Heckman 1990; Bomans \etal 1997; Franx \etal 1997; Martin 1999; Pettini \etal 2001; Frye, Broadhurst, \& Benitez 2002; Rupke \etal 2005; Veilleux \etal 2005; Weiner \etal 2009; Martin \etal 2013).  They are thought to cause the strong correlation between mass and metallicity observed in low-mass galaxies (\eg Dekel \& Silk 1986; Tremonti \etal 2004; Erb \etal 2006; Kewley \& Ellison 2008); they are needed to reconcile the number density of observed galaxies with the favored cosmological model (e.g. Somerville \& Primack 1999; Cole \etal 2000; Scannapieco \etal 2002; Benson \etal 2003); and they are essential to the enrichment of the intergalactic medium (Tytler \etal 1995; Songaila \& Cowie 1996; Rauch \etal1997; Simcoe \etal 2002; Pichon \etal 2003; Schaye \etal 2003; Ferrara \etal 2005; Adelberger \etal 2005;  2006; Steidel \etal 2010; Martin \etal 2010). Yet, despite the central importance of galaxy outflows, the processes that control their evolution are extremely difficult to constrain both theoretically and observationally. 

From a theoretical point of view, the most uncertain issue is the coupling of stars to the surrounding interstellar medium (ISM). This is because the highly efficient cooling within the ISM makes it impossible to model supernovae by adding thermal energy to the medium, while, at the same time, the range of physical scales involved does not allow for the direct modeling of supernovae within a galaxy-scale simulation. As a result, studies have been forced to adopt a number of unsatisfactory approximations, including: temporarily lowering the densities of heated particles or delaying their cooling (\eg Gerritsen \& Icke 1997; Thacker \& Couchman 2000; Stinson \etal 2006),  imposing a minimum temperature floor (Suchkov \etal  1994; Tenorio-Tagle \& Mu–›oz-Tu–—\~ non 1998; Strickland \& Stevens 2000; Fujita \etal 2004), using an empirical heating function to mitigate cooling (Mac Low \etal 1989; Mac Low \& Ferrara 1999), implementing exaggerated momentum kicks (Navarro \& White 1993; Mihos \& Hernquist 1994; Scannapieco \etal 2001), and temporarily decoupling particles from their neighbors (Springel \& Hernquist 2003; Scannapieco \etal 2006; Dalla Vecchia \& Schaye 2008). In fact, even the most detailed cosmological `zoom-in' simulations  (\eg Gnedin \etal 2009; Agertz \etal 2009; Ceverino \etal 2010; Governato \etal 2010; Shen \etal 2012) are faced with the problem that excessive cooling is exacerbated by the fact that supernovae often go off within giant molecular clouds that must be pre-conditioned by ionization fronts (\eg Matzner 2002) and radiation pressure (\eg Murray \etal 2010) to be modeled accurately. Although recent efforts have attempted to couple stars with the ISM in more realistic detail (\eg Hopkins \etal 2011, 2012a,b), even these rely strongly on tuning against observations to achieve good results.

From the observational point of view, the most important issue is measuring and interpreting the evolution of the wide range of multiphase material found in galaxy outflows.
This ranges from $\approx10^7-10^8$K plasma observed in X-rays (\eg Martin \etal 1999; Strickland \& Heckman 2007, 2009), to $\approx10^4 $K material observed at optical and near UV wavelengths  (\eg Pettini 2001; Tremonti \etal 2007; Martin 2012; Soto \& Martin 2012), to $10-10^3$ K molecular gas observed at radio wavelengths (\eg Walter 2002; Sturm 2011; Bolatto \etal 2013).   Furthermore, the easiest phase to interpret, the X-ray emitting medium, is the most difficult to observe. In fact, the $10^8$ K medium is so hot that it is only detectable in deep Chandra and XMM imaging in very nearby galaxies (\eg Strickland \& Heckman 2007; Wang \etal 2014) and best measured in M82, where it appears to be well fit by a simple analytic model (Chevalier \& Clegg 1985; Heckman \etal 1990).  Below $10^7$ K, the X-ray emitting medium is detectable in a larger number of galaxies, and it appears to be best understood in terms of ambient material that is shock heated by the wind fluid as it flows out to large distances (\eg Suchkov \etal 1994; Strickland \& Stevens 2000). Given the high temperatures of the X-ray emitting plasma, it will always escape the gravitational potential of its host, and its temperature, surface brightness, and other properties are remarkably regular over a wide variety of galaxies (Grimes \etal 2005).

The colder phases, on the other hand, can be easily observed from the ground and studied at many redshifts, but they are  poorly understood theoretically.  Unlike the X-ray emitting medium, low-ionization state material is observed to have complex velocity profiles (\eg Westmoquette \etal 2009; 2012) that are strongly correlated with the overall host luminosity (\eg Martin 2012) circular velocity (\eg Martin 2005), star formation rate per unit area (\eg Chen \etal 2010) and star formation rate per unit stellar mass (\eg Heckman 2014). However, because it is often visible only through absorption lines and resonant Lyman-alpha emission (Pettini \etal 2001), the total mass in this phase is poorly constrained.  It could either be the primary avenue for baryon and metal ejection, or make up only a small fraction of the ejected material. Equally poorly understood is the final fate of this material, as its position along the line of sight is unknown and it is often moving at velocities that are similar to the escape velocity of the host. In fact, even the presence of this medium is surprising, as simple theoretical estimates predict that it should be disrupted by interactions with the hot wind fluid well before it is accelerated to significant velocities (\eg  Scannapieco 2013). 

Furthermore, this uncertainty is somewhat surprising given that simulations of cold clouds interacting with a hot, high-velocity medium have  been carried out by many authors, with simulations going back almost 50 years.   Groups have carried out both two and three-dimensional simulations: neglecting radiative cooling (Nittmann \etal 1982; Klein \etal 1994), including radiative cooling (Woodward 1976; Mellema \etal 2002; Fragile \etal 2004; Melioli \etal 2005; Cooper \etal 2009; Marinacci \etal 2010, 2011), including both radiative cooling and thermal conduction (Marcolini \etal 2005; Orlando \etal 2005, 2006, 2008; Recchi \& Hensler 2007), including the impact of magnetic fields  (Mac Low \etal 1994; Gregori  \etal 1999, 2000; Fragile \etal 2005; Orlando \etal 2008; Shin \etal 2008), and including nonequillibrium chemistry effects (Kwak \etal 2011; Henley \etal 2012). Yet, despite the usefulness of these studies in elucidating the physics of cold clouds in a hot medium, none of them has spanned the range of parameters and timescales necessary to study galaxy outflows.

The key issue in this case is the ability for the hot wind to accelerate cold material, before it disrupts it. In the definitive study of the non-radiating, hydrodynamic case, Klein \etal (1994) showed that if hot material moves past a cloud at a velocity $v_{\rm hot}$ that is much greater than the cloud's internal sound speed,  it will shred the cloud on a `cloud crushing' timescale 
\be
t_{\rm cc} \equiv \frac{\chi_0^{1/2} R_{\rm cloud}}{v_{\rm hot}},
\ee where  $R_{\rm cloud}$ is the cloud radius and $v_{\rm hot}/\chi_0^{1/2}$ is the velocity that the resulting shock moves through the cloud, with $\chi_0$ the initial density ratio between the cloud and the surrounding medium.   Subsequent studies showed that magnetic fields and radiative cooling can delay this disruption by a few cloud crushing times but not prevent it
(\eg Mac Low \etal 1994; Orlando \etal 2005, 2008). On the other hand, to accelerate the cloud to the hot wind velocity, the impinging mass must be comparable to the cloud mass, which as discussed in more detail below, takes a time $t_{\rm accel} \approx  (4 \pi/3) R_{\rm cloud}^3 /[ \pi R_{\rm cloud}^2 v_{\rm hot}] \approx \chi_0^{1/2} t_{\rm cc}.$ Because the  $\approx 10^4$ K clouds observed in galaxy outflows are in rough pressure equilibrium  with the surrounding medium (Strickland \& Heckman 2007),  $\chi_0 = T_{\rm hot} /10^4 {\rm K}$ = $300-10^4$ such that $t_{\rm accel} \geq 20 t_{\rm cc}$ (\eg Scannapieco 2013). Thus it would appear that cold clouds never survive to reach velocities comparable to that of the hot wind.

There is, however, one possible caveat. If cloud cooling is efficient and $v_{\rm hot}$ exceeds the sound speed in the hot medium, $c_{s,{\rm hot}},$ a bow shock develops in front of the cloud, which protects it from ablation, both by reducing heating and squeezing it to even higher densities.   When combined with radiative cooling, these mitigating effects allow the cloud to remain intact for many cloud crushing times.  In fact, the acceleration of clouds  when $M_{\rm hot} = v_{\rm hot}/c_{s,{\rm hot}}$ is large has been simulated by Cooper \etal (2009) and Kwak \etal  (2011), but  in both studies the cloud was accelerated out of the simulation volume before a final answer was obtained.

In this series of papers, we overcome this limitation by carrying out adaptive mesh refinement (AMR) simulations in which we automatically shift the frame of the calculation to the center of mass frame of the cloud once per cloud crushing time, allowing us to run until the cloud is fully disrupted, regardless of the exterior Mach number.   In this way, we are able to carry out a suite of simulations that spans the parameter space relevant for hot material interacting with the cold clouds that provide us with the bulk of the observational constraints on galaxy outflows.   In this first paper, we study both cloud mass loss and cloud acceleration accounting for radiative cooling.  Future papers will study: (i) the impact of electron thermal conduction (\eg Cowie \& McKee 1977), which can provide an additional source of cloud heating, leading to more rapid disruption, and (ii) the impact of magnetic fields, which can extend the lifetime of the cloud, both by suppressing thermal conduction perpendicular to the field lines (\eg Braginskii 1965; Bogdanovi{\'c} \etal 2009; Parrish \etal 2009) and stabilizing the KH instability (\eg Chandrasekhar 1961; Walker 1981; Jun \etal 1995; Br{\"u}ggen \& Hillebrandt 2001; Dursi \& Pfrommer 2008).

The structure of this paper is as follows.  In Section 2, we describe the physics of cold-cloud hot-wind interactions in more detail, outlining the key parameter space. In Section 3, we describe our numerical methods including changes of frame and selective de-refinement criteria that allow us to obtain results out to many cloud crushing times. In Section 4, we present our simulation results,  give fitting formulae for the mass and velocity of the clouds as a function of time  and exterior conditions, and we discuss the underlying physics that leads to these scalings. Conclusions are given in Section 5.

\section{Physics of Cold Cloud Driving by Hot Galaxy Outflows}

\subsection{Hot Wind Properties}

We are interested in clouds impacted by a hot wind expanding rapidly from a starbursting galaxy. For most galaxies, this hot medium moves many scale heights per Myr, while a typical starburst episode  lasts for many Myrs (\eg Greggio \etal 1998; F\"orster Schreiber \etal 2003). Thus, the distribution is expected to be well approximated by an equilibrium configuration, and this seems to be born out observationally in most objects in which reliable X-ray analyses can be made (\eg Heckman 1990; Heckman \etal 1995; Ott \etal 2005; Strickland \& Heckman 2007; Yukita \etal 2012).
In this case, assuming the expansion is spherically symmetric (or equivalently, assuming a conical expansion with a fixed opening angle)
 the equations of mass, momentum, and energy conservation are
\be
\frac{1}{r^2} \frac{d}{dr} (\rho_{\rm hot} v_{\rm hot} r^2) = \dot q_{\rm m},
\ee
\be
\rho v_{\rm hot} \frac{dv_{\rm hot}}{dr} = - \frac{dP_{\rm hot}}{dr} - \dot q_{\rm m} v_{\rm hot},
\label{eq:mom}
 \ee
\be
\frac{1}{r^2} \frac{d}{dr} \left[ \rho_{\rm hot} v_{\rm hot} r^2 \left( \frac{v_{\rm hot}^2}{2} + \frac{\gamma}{\gamma -1} \frac{P_{\rm hot}}{\rho_{\rm hot}} \right) \right] =  \dot q_{\rm e},
\ee
where $\rho_{\rm hot}$, $v_{\rm hot}$, $P_{\rm hot},$ and $\gamma$ are the density, radial velocity,  and ratio of specific heats of the hot medium,
and the mass and energy input rate are
\be
\dot q_{\rm m} =  
\begin{cases}
\dot q_{\rm m,0} & \text{if } r \leq R_\star \\
   0       & \text{if } r > 0,
\end{cases}
\ee
and $\dot q_{\rm e} =  \dot q_{\rm m}c_{s,{\rm hot},0}^2/ (\gamma-1),$ respectively, where
$R_\star$ is the driving radius of the flow and
$c_{s,{\rm hot},0} = [(\gamma-1) \dot q_{\rm e,0}/\dot q_{\rm m,0} ] ^{1/2}$ is the sound speed of the hot medium at $r =0.$  For reference, in the case of M82, $R_\star \approx 300$ pc (Strickland \& Heckman 2009).

The solution to these equations is the Chevalier and Clegg (1985) model, which can be used to gain a good understanding of the hot wind conditions as a function of radius.  For a $\gamma = 5/3$ gas, the sound speed  when $r < R_\star$ is approximately constant at 
\be
c_{s,{\rm hot}}(r) \approx c_{s,{\rm hot},0} = 0.82 \left( \frac{\epsilon \dot E}{ \beta \dot m} \right)^{1/2}, 
\ee
where $\dot E$ is the total energy input from supernovae per unit time, $\epsilon$ is the fraction of this energy that is deposited into the hot medium, $\dot m$ is the total mass ejected by supernovae per unit time, and $\beta$ is the mass input per unit time into the hot material, which may exceed one due to entrainement. The outward velocity of the medium, on the other hand, increases approximately linearly in the driving region as $v_{\rm hot}(r) = c_{s,{\rm hot},0} r/4 R_\star$, such that the Mach number increases as $M_{\rm hot}(r) \approx  r/4 R_\star$,  where here and below we use $M$ to distinguish Mach number from mass, which is always denoted as $m$.

At the edge of the driving region, the solution reaches the sonic point, with  $v_{\rm hot}(R_\star) = c_{s,{\rm hot}}(R_\star) = (\epsilon \dot E/ 2 \beta \dot m)^{1/2}$.  Finally, outside the driving region,  the radial velocity quickly approaches a constant value of $v_{\rm hot}(r) = (2 \epsilon \dot E/ \beta \dot m)^{1/2} $ and the Mach number goes as $M_{\rm hot}(r) = 2^{5/3}  (r/R_\star)^{2/3},$ such that $c_{s,{\rm hot}}(r) = 2^{-5/3}  (2 \epsilon \dot E/ \beta \dot m)^{1/2}  (r/R_\star)^{-2/3}.$      Note that at all radii, the Mach number is purely a function of $r/R_\star$ and independent of the rate of mass and energy input into the hot wind.

Thus  sampling a series of Mach numbers with a suite of simulations corresponds to  sampling the properties of the hot wind as a function of radius.  These distances have  been tabulated in column 3 of Table 1 for a range of Mach numbers.  Note that the velocity and sound speed change rapidly near the sonic point, such that $M_{\rm hot}$ goes from 0.5 to 3.5 between $0.9$ and $1.1$ $r/R_\star.$    The  Chevalier and Clegg (1985) model also gives an estimate of the density of the expanding medium, which is indicated  in column 4 of Table 1 as $n/n_0$,  the number density of the medium relative its number density at $r=0.$  Again, for reference, the pressure of the $\approx 10^8$K medium  in  M82  is $P / k \approx 1-3 \times 10^7$ cm$^{-3}$ K  in the driving region, which corresponds to  $n_0 \approx 0.1-0.3$ cm$^{-3}.$ 

If we take a standard estimate of $\dot E/\dot m$ of  $10^{51} {\rm ergs } / (10 M_\odot )$, this gives a velocity $(2 \epsilon \dot E/\beta \dot m)^{1/2}= (\epsilon/\beta)^{1/2} \, 3160$ km s$^{-1}$ at large radii and a temperature at small radii of  $ (\epsilon/\beta) \, 1.1 \times 10^8$ K or 9.3 keV. This is only slightly higher than observed in M82 (Strickland \& Heckman 2009), implying only moderate mass loading is typical for many starbursts, with   $\epsilon/\beta$, usually $\geq 0.3$ and always  $\geq 0.1.$   This gives us a reasonably small parameter space of conditions of the exterior flow that are most important for understanding cold-cloud acceleration.

Table 1 also shows $v_{\rm hot}$ and $T_{\rm hot}$ for a number of choices of $\epsilon/\beta,$ for the cases described in more detail below.   These values were chosen to span a wide range of efficiencies, focus on the poorly-studied $M_{\rm hot} \geq 1$ regime, and provide multiple runs with the same temperature and/or velocities, such that we can study the impact of changing $M_{\rm hot}$ while maintaining $v_{\rm hot}$ or $c_{s,{\rm hot},0}$ constant.  Thus we only consider a single choice of  $\epsilon/\beta$ for $M_{\rm hot}=0.5$ case, but we consider three choices for $M_{\rm hot}=1$: one that gives a similar $v_{\rm hot}$ as the $M_{\rm hot}=0.5$ case, one that gives the same $c_{s,{\rm hot}}$ as the $M_{\rm hot}=0.5$ case, and one for which $\epsilon/\beta \approx 1.$ Similarly, for $M_{\rm hot} \approx 3.5$ we adopt three choices of $\epsilon/\beta$ that allow for comparisons with the $M_{\rm hot}=1$ cases with the same $c_{s,{\rm hot},0}$ and similar $v_{\rm hot}$ values. Note that in one case, this forces us to take $\epsilon/\beta$ slightly greater than 1, which is not unreasonable given the uncertainties in the mass and energy input from supernovae. Finally, in the highest three mach number cases we chose $\epsilon/\beta$ to exactly match the $c_{s,{\rm hot},0}$ and $v_{\rm hot}$ taken in the $M_{\rm hot} \approx 3$ cases.  Together these choices  allow us to define a small number of runs that can nevertheless be used to understand the evolution of cold clouds in a wide range of outflow conditions.

\begin{table*}
\caption{Simulation Parameters}
\begin{centering}
\resizebox{\textwidth}{!}{%
\begin{tabular}{|l|rcccccccccrc|}
\hline
Name & $M_{\rm hot}$ & $r/R_\star$ & $n/n_0$ &  $\epsilon/\beta$ & $v_{\rm hot}$               &    $T_{\rm hot}$         & $T_{\rm hot}$ &    $\chi_0$ &  $M_{\rm cloud}$ & $t_{\rm cc}$ &$N_{\rm cool}$  & $\Sigma_{\rm cool}$  \\
           &        		    &  		       & 		   &				& km s$^{-1}$			 &  $10^6$ K      	&  keV 		&                    &    		  &  Myr/100pc   &  cm$^{-2}$  	 	& $M_\odot {\rm pc}^{-2}$  \\
   \hline
  M0.5v430     &    0.5  & 0.9     &  0.8 & 0.2 & 430   & 30      &  2.7        &     3000  & 28 & 12.5    		&  $10^{17}$ 		&  $5 \times 10^{-4}$  \\ 
  M1v480        &    1.0  &  1.0    &   0.4    & 0.1 & 480   & 10   & 0.86         &     1000 & 32  	& 6.4 &   $10^{17}$		& $5 \times 10^{-4}$ \\ 
  M1v860         &    1.0 &  1.0    &   0.4   &  0.3  & 860   & 30   & 2.7     &      3000   & 57    	&6.2 &   $10^{17}$		&$5 \times 10^{-4}$   \\
  M1v1500      &    1.0  &   1.0   &    0.4  & 0.9   &1500 & 100 & 8.6   &      10000  & 100 & 6.5  		& $10^{17}$ 		& $5 \times 10^{-4}$  \\
  M3.8v1000   &    3.8  &  1.1    &   0.2   & 0.1   & 1000 &   3   & 0.27    &      300   &  66  & 1.7     		&  $10^{17.5}$	 & $1.5 \times 10^{-3}$  \\
  M3.5v1700   &    3.5  &   1.1   &   0.2   & 0.4  &1700 &   10  & 0.86  &     1000  & 110  & 1.8   		&  $10^{17.5}$ & $1.5 \times 10^{-3}$  \\
  M3.6v3000   &    3.6  &   1.1   &   0.2   & 1.1   & 3000 &   30 & 2.7     &      3000  &  200 & 1.8    		&  $10^{17.5}$ &$1.5 \times 10^{-3}$  \\
  M6.5v1700   &    6.5  &   1.9   &   0.05 & 0.3   & 1700 &   3   & 0.27    &       300   & 110  & 1.0 		& $10^{18}$ 		&$5 \times 10^{-3}$  \\
  M6.2v3000   &    6.2  &   1.9   &  0.05  &   1.0  & 3000 &  10  & 0.86  &   1000  & 200 & 1.0  		& $10^{18}$ 		& $5 \times 10^{-3}$  \\
  M11.4v3000  &   11.4 &   2.6  &   0.03  &  0.9 & 3000 &  3    & 0.27    &      300  &  200  & 0.56   		& $10^{19}$ 		&$5 \times 10^{-2}$  \\
\hline
\end{tabular}}
\end{centering}
\\
\end{table*}

\subsection{Cold Cloud Disruption}

The clouds of interest will have temperatures $\approx 10^4$ K, due to the balance between photoheating by the ionizing background and strong radiative cooling between $10^4$ and $10^{5.5}$ K.    For a medium at $10^4$K,  the Jeans length, $\lambda_J  \approx c_{\rm s, cloud} (G \rho)^{-1/2} \approx 2 \, {\rm kpc} \, (n/{\rm cm}^3)^{-1/2}$, meaning that the clouds will be pressure-confined rather than gravitationally bound.  For such clouds, the initial ratio of the cloud density to the density of the exterior medium, $\chi_0,$ will also be equal to the ratio of the exterior temperature to the cloud temperature. This allows us to associate a single density contrast with each exterior temperature, as shown in column 9 of Table 1.  The fixed temperature of the clouds also allows us to compute a single cloud Mach number $M_{\rm cloud}= v_{\rm hot}/c_{\rm s,cloud},$ shown in column 10 of this table.

As the sound crossing time for these pressure confined clouds is only $\approx  0.05 \, {\rm Myr} \, R_{\rm cloud}/{\rm parsec}$, we can reasonably approximate them as spherical before they encounter the shock.    This assumption also makes the simulations of their evolution easier to interpret, as well as allows us to make closer contact with the rich existing literature on the topic.  There are a few important timescales that determine the evolution of cold clouds in this case.  The most important of these is the cloud crushing time mentioned above, $t_{\rm cc}  = R_{\rm cloud}/(v_{\rm hot} \chi_0^{1/2}).$  In the Appendix, we relate $v_{\rm hot},$ the velocity of the wind impacting the cloud to $v_{\rm t},$ the velocity of the initial shock transmitted through the cloud, and show that in the limit of high Mach numbers $v_t \approx v_{\rm hot}/\chi_0^{1/2}.$   Thus the cloud crushing time gives a rough estimate of when the shock from the exterior medium moves through the cloud, heating it and disrupting it if it is not able to radiate the energy away efficiently.  In column 11 of Table 11 we show the value of this time in units of Myrs for clouds of size $R_{\rm cloud} = 100$pc.   

A second important timescale is the cooling time behind the transmitted shock, which can be estimated as 
\be
 t_{\rm cool} = \frac{3/2 n_{\rm cloud} kT_{\rm t}}{\Lambda(T_{\rm t}) n_{\rm e,cloud} n_{\rm i,cloud}}, 
\ee
where $k$ is the Boltzmann constant, $T_{\rm t}$ is the post-shock temperature, $\Lambda(T_{\rm t})$  is the equilibrium cooling function evaluated at $T_{\rm t}$, and $n_{\rm cloud},$ $n_{\rm e,cloud},$ and $n_{i\rm ,cloud},$   are the total, electron, and ion number densities within the cloud, respectively.
The ratio of this time to the cloud crushing time is $t_{\rm cool}/t_{\rm cc} =  N_{\rm cool}/(n_{\rm i,cloud} r_c),$
where the column density, $N_{\rm cool} \equiv  3 k T_{\rm t} v n_{\rm cloud} [ 2 \Lambda \chi^{1/2} n_{\rm e,cloud} ]^{-1}$ 
is purely a function of the velocity of the transmitted shock.

Because the clouds of interest are at a fixed temperature and the Mach number of the transmitted shock within the cloud can be related to the Mach number in the exterior medium, this means that $N_{\rm cool}$ can be well approximated by a function of $M_{\rm hot}.$   These numbers have all been added to Table 1, as calculated using the equilibrium cooling curves in Wiersma \etal (2009), assuming solar metallicity and an average atomic mass of $0.6.$    These values indicate that for the conditions of interest, the cooling time will be much shorter that the cloud crushing time for all but the smallest, sub-parsec size clouds, and thus radiative cooling will be very efficient throughout the evolution of most observed clouds. 

A third important timescale is  the characteristic timescale for disruption by the shear-driven Kelvin-Helmholtz (KH) instability.  In a linear stability analysis, the width of the KH layer grows at a rate $\Delta v_{\rm KH} \propto \Delta v \chi^{-1/2},$ where $\Delta v$ is the velocity difference between the cloud and the hot medium (Chandresekshar 1961).   
Similarly, experimental measurements of subsonic shearing flows show that the nonlinear growth across the boundary with a high density ratio is asymmetric, such that the ``entrainment ratio'' of the width on the high density side of the layer to the width on the low density side of the layer  $E_v = \chi^{-1/2}$  (Brown 1974; Konrad 1976).  
 
This  can be understood in terms of entrainment into a spatially-growing shear layer made up of large-scale vortical structures convecting at a velocity $v_c$ (Coles 1981; Dimotakis 1986).  In the subsonic case, $v_c$ is set by the condition that in the frame of the vortices
\be
P_1 +  \rho_1 (v_1 - v_c)^2 \approx \rho_2 (v_2 - v_c)^2  + P_2,
\label{eq:pressures}
\ee
where $P_1$ and $P_2$ are the pressures of the material on either side of the layer, $\rho_1$ and $\rho_2$ are their densities and
$v_1$ and $v_2$ are their velocities.   This means that if $P_1 = P_2$, then $| v_1-v_c | = \chi^{1/2} | v_2 - v_c|$, such that the heavy material drags the vortices along with it.   The expansion of the shear layer on either side of the density contrast in this case appears to be roughly proportional to  $|v - v_c |$  (Brown \& Roshko 1974; Papamoschou \& Roshko 1988; Slessor \etal 2000), leading to a strongly asymmetric growth of the boundary with the width on the dense side as a function of time given by
\be
\delta \approx 0.1 t \Delta v \chi^{-1/2}.
\ee
Thus, even if cooling is efficient, the subsonic KH instability will grow to the scale of the cloud within a few cloud crushing times.   

In the supersonic case, however, the growth of the boundary layer is much slower (Chinzei \etal1986; Papamoschou \& Roshko 1988; Samini \& Elliott 1990;  Goebel \& Dutton1991; Hall et al 1993; Barre \etal 1994; Clemens \& Mungal 1992; Naughton \etal 1997; Slessor \etal 2000), but the level of this suppression, and the structure of the mixing layer are much more poorly understood.
Papamoschou \& Roshko (1988) measured this suppression as a function of convective Mach numbers:
\ba
M_{c,1} \equiv \frac{v_1 - v_c}{c_{s,1}} \qquad {\rm and} \qquad M_{c,2} \equiv \frac{v_c - v_2}{c_{s,2}},
\ea
where again $v_c$ is the convection velocity of the turbulent structures and $c_{s,1}$ and $c_{s,2}$ are the sound speeds of the two media.
Although in this case the best choice for $v_c$ is unclear.   If one adopts the isentropic pressure-recovery model of Papamoscho \& Roshko (1988), then if $\gamma_1=\gamma_2$,  $v_c$ is the same as given by eq.\ (\ref{eq:pressures}) and $M_{c,1} = M_{c,2} = \Delta v/(c_{s,1}+c_{s,2}).$ As a function of this quantity, the width of the layer is reduced by a factor of  $\delta(M_c)/\delta(0)$ which approaches $1/4$, as $M_c$ approaches the largest experimentally measured values $\approx 1.5.$
 
On the other hand, several authors have argued for different choices for $v_c$, such as the phase speed of the linearly most unstable mode
(Ragab \& Wu 1989), the phase speed of the linearly most unstable mode at a point corresponding to neutral stability (Sandham \& Reynolds 1989), and the speed of turbulent structures in the presence of asymmetric shocks (Dimotakis 1991).  In fact, experimentally it appears that supersonic shear layers favor a configuration in which $M_{c,1}$ and $M_{c,2}$ are very different, such only one side of the layer is supersonic (Papamouschou 1991). With this asymmetry in mind, Slessor et al (2000) proposed a suppression of the width of the KH layer by a factor 
\be
\delta(M_c)/\delta(0) \approx  [1 +4 (\gamma-1) M_{\rm KH}^2]^{-1/2},
\label{eq:Slessor}
\ee
where $M_{\rm KH}=\Delta v/c_{\rm s,min}$ is the ratio of $\Delta v$ with $c_{\rm s,min},$ the {\em minimum} of the two sound speeds of the fluids that are being sheared (Slessor \etal 2000).   Extrapolating this fit formula to the cases of interest reduces the growth of the shear layer by an even large factor than the Papamoschou \& Roshko (1988) scaling, with $\delta(M_{\rm KH})/\delta(0) \approx M_{s,{\rm cloud}}$ at very large Mach numbers.

\subsection{Velocity Evolution}

The radial velocity of the cloud, $v_{\rm cloud},$ will depend on the momentum imparted by the hot wind, divided by the total cloud mass.
For a spherical cloud this gives
\ba
v_{\rm cloud} (t)  &= & \int_0^t dt' \frac{\pi R_{\rm cloud}(t')^2 \Delta v(t')^2}{\frac{4 \pi}{3} R_{\rm cloud}(t')^3 \chi(t') } \nonumber \\
&=&  \frac{3 v_{\rm hot} }{4 \chi^{1/2}_0 } \int_0^{t/t_{\rm cc}}  d \tilde t'  \, \tilde R_{\perp}^2 (\tilde t') \Delta \tilde v(\tilde t')^2,
\label{eq:vcoft}
\ea
where $\tilde t$ is the time in units of the initial cloud crushing time, $\chi_0$ is again the initial density contrast, and $\Delta \tilde v$ and $\tilde R_{\perp}$  are
the relative velocity between the cloud of the exterior medium and the radius of the cloud perpendicular to the flow in units of their initial values, respectively. Notice that because this ratio compares the cross section the cloud presents to the incoming flow with the total mass of the cloud, the final expression 
does not depend on the size of the cloud in the direction of the incoming material. Notice also that if we consider the evolution of the cloud in units of the cloud crushing time, the physical cloud radius does not appear in this expression, only $\tilde R_{\perp} =  R_{\perp} /R_{\rm cloud}.$

Eq.\ (\ref{eq:vcoft}) also illustrates the fact that it is more difficult to accelerate clouds with larger density contrasts, although this effect is somewhat mitigated  because the cloud is initially flattened by the collision (\eg Klein \etal 1984).  In the limit in which $\Delta \tilde v$,  and $\tilde R_{\perp}$  are fixed, the distance traveled by the cloud is proportional to the square of its lifetime in units of the cloud crushing time.   This makes the results especially sensitive to the late time evolution of the cloud, and it also means that the suppression of the KH instability can be very important in determining the distance the cloud travels before it is disrupted.

\section{Methods}

\subsection{Setup}

To study cold-cloud hot-flow interactions over the range of conditions encountered in galaxy outflows, we carried out a suite of simulations using  FLASH (version 4.2), a multidimensional hydrodynamics code (Fryxell \etal  2000) that solves the Riemann problem on a Cartesian grid.   Each of these runs adopted one  of the sets of representative conditions described in Table 1.  All simulations were three-dimensional, as imposing a 2D cylindrical geometry reduces the degrees of freedom over which shear-driven instabilities can develop, which is likely to have a large impact on the evolution of the cloud (\eg Pan \etal 2012).   

In all cases, we used the default directionally-split  (Strang 1968) Piecewise-Parabolic Method hydrodynamic solver (PPM; Colella \& Woodward 1984; Colella \& Glaz 1985; Fryxell, M\" uller, \& Arnett 1989).  We chose this approach as we found that it was much better at preserving spherical symmetry in test simulations in which a cloud was compressed by a stationary high-pressure medium.   We also made use of the shock detect flag which lowered the prefactor in the Courant-Friedrichs-Lewy  timestep condition from its default value of 0.4 to 0.25 in the presence of strong shocks.

Because of the scalings discussed in \S2, as long as the column density is sufficiently large that cooling is efficient, the size of the cloud scales out of the problem if we express our results in units of the cloud crushing time.   Thus without a loss of generality, we chose a fixed cloud radius of 100 parsecs for all our simulations, an initial temperature of $10^4$K, and a mean density of $\rho = 10^{-24}$ g cm$^{-3},$ such that   $\rho R_{\rm cloud} = 1.4 M_\odot$ pc$^{-2},$ and $n_{\rm i,cloud} R_{\rm cloud} = 3.1 \times 10^{20}$ cm$^{-2}$.  In all simulations,  the computational domain covered a physical volume of $-800 \times 800$ parsecs in the $x$ and $y$ and directions and $-400$ to $800$ parsecs in $z$ direction, where the cloud was initially centered at (0,0,0) and $z$ is the direction of the hot outflowing material.  Outside of the cloud, the initial velocity and the sound speed of the material were taken to be $v_{\rm hot}$, and $c_{s,{\rm hot}}$ as given by Table 1, and the density was set by pressure equilibrium with the cold cloud.
 
 At the lower $z$ boundary we continuously added material to the grid with the same values of $v_{\rm hot},$ $c_{s,{\rm hot}},$ and density as in our initial conditions.  In the $x$ and $y$ directions, as well as at the $+z$ boundary, we adopted the FLASH ``diode" boundary condition, which assumes a zero normal gradient for all flow variables except pressure and does not allow material to flow back onto the grid.  The large $x$ and $y$ size of the simulation was chosen such that this zero gradient condition did not effect the shape of the shock, which can become unnaturally planar in smaller simulation domains.   Likewise a significant standoff distance was left in front of the cloud, such that the front of the bow shock remained within the simulation volume, and a significant distance was left behind the cloud, to capture the evolution of the disrupted material.

 \subsection{Cooling, Frame Changing, and Refinement/De-refinement Criteria}
 
Cooling was computed in the optically-thin limit,  assuming local thermodynamic equilibrium as   
\be
\dot E_{\rm cool}  =  (1-Y) \left(1- \frac{Y}{2} \right) \frac{\rho \Lambda}{(\mu m_p)^2},
\label{eq:ecool}
\ee 
where $\dot E_{\rm cool}$ is the radiated energy per unit mass, $\rho$ is the density in the cell, $m_p$ is the proton mass, $Y=0.24$ is the helium mass fraction, $\mu=0.6$ the mean atomic mass, and $\Lambda(T,Z)$ is the cooling rate as a function of temperature and metallicity.     Here we made use of the tables compiled by  Wiersma \etal (2009) from the CLOUDY code (Ferland \etal 1998), making the simplifying approximations that the metallicity of the material is always solar and that the abundance ratios of the metals always occurs in solar proportions.   As in Gray \& Scannapieco (2010), subcycling was implemented  within the cooling routine itself, such that $T$ and $\Lambda(T,Z)$  were recalculated every time $E_{\rm cool}/E > 0.1$.  This is equivalent to an integral formalism that assumes a constant density over each hydrodynamic time step (\eg Thomas \& Couchman 1992; Scannapieco, Thacker, \& Davis 2001). We did not include heating by a photoionizing background in our calculations, but the approximate impact of including this would be to slightly raise the minimum temperature below which the cloud can cool effectively, while having a minor effect at higher temperatures.   Likewise changing the metallicity would lengthen the cooling times somewhat, particularly above $10^{4.5}$K, but not change the overall conclusion of the efficiency of cooling for the majority of the clouds of interest.
 
In order to be able to run our simulations until the clouds were disrupted, while still keeping them on the grid, we developed an automated frame changing routine.  By labeling the cloud material with a scalar, we were able to track its center of mass position  and velocity, ${\bf x_{\rm cloud}}$ and ${\bf v_{\rm cloud}}$, as well as its radial extent in the $x$, $y$, and $z$ directions,  calculated as the mass weighted average values of ${\rm abs}(x-x_{\rm cloud}),$ ${\rm abs}(y-y_{\rm cloud}),$ and ${\rm abs}(z-z_{\rm cloud}).$ Every cloud crushing time, starting at $t=3 t_{\rm cc},$ we checked if the cloud was moving in the positive $z$ direction in the frame of the simulation and if $z_{\rm cloud} \geq R_{\rm cloud}/2,$  such that the cloud was not too near the front of the simulation volume.   If these criteria were satisfied, we shifted every $z$ velocity in the simulation by a constant factor of 1.2 times the instantaneous cloud velocity, appropriately adjusting the kinetic energy.  This moved the simulation from a frame in which the cloud was rapidly drifting towards the $+z$ boundary, to a new frame in which the cloud was slowly drifting towards the $-z$ boundary.  As more momentum accumulated on the front of the cloud, it would once again start to move toward the $+z$ boundary, until another cloud crushing time went by and the frame was again shifted to make the cloud slowly drift upstream. In this way, we were able to keep the cloud center of mass near $z=0$ at all times, and still reconstruct its evolution in the frame of the wind by keeping track of the overall frame shift.

To determine when zones are refined and de-refined, FLASH uses the second derivatives of ``refinement variables," normalized by their average gradient over a cell. In the default FLASH configuration, if this number is greater than 0.8, the cell is marked for refinement, and if all the cells in a block lie below 0.2, they are marked for de-refinement.   We used both density and temperature as refinement variables in this way, but also adopted a set of additional refinement and de-refinement criteria, chosen to minimize the computational cost of the simulation while at the same time maintaining the most accurate results possible in the spatial regions that are the most important to the evolution of the cold cloud.

While the large simulation volume perpendicular to the $z$ axis was needed to maintain the proper angle for the Mach cone that forms around the cloud, it is not important for us to achieve high resolution for this purpose.   Thus we forced our simulation to automatically mark cells for de-refinement if either of two criteria were satisfied: (i) if the distance from a cell to the $z$ axis was greater than three times the original cloud radius or nine times the instantaneous $x$ extent of the cloud or if  (ii) the distance  from a cell to the $z$ axis was greater than either the original cloud radius or three times the instantaneous $x$ extent of the cloud {\em and also} both ${\rm abs}(z)$ and ${\rm abs}(z-z_{\rm cloud})$ were greater than 3 $R_{\rm cloud}.$ 

Likewise, in order to obtain the most reliable results possible, it is important for us to maintain high resolution consistently in the regions within the cloud and the shear layer immediately around it.  To achieve this in the initial stages of the interaction, when $t \leq 2 t_{\rm cc},$ we automatically marked cells for maximum refinement within the cylindrical region with ${\rm abs}( z) \leq 1.5 R_{\rm cloud}$ and the distance from the $z$ axis $(x^2+y^2)^{1/2} \leq 1.5 R_{\rm cloud}.$ Once the transmitted shock made its way all the way through the cloud at  $t > 2 t_{\rm cc},$ we then only forced the simulation to maintain maximum refinement when ${\rm abs} (z-z_{\rm cloud}) \leq R_{\rm cloud}$ and $(x^2+y^2)^{1/2} \leq 1.5 R_{\rm cloud},$ although, in practice the complex density and pressure structure that developed around the cloud by this time led to highly-refined regions that occupied a much larger volume, approaching the full volume in which derefinement was not forced by our criteria above.

\begin{figure*}
\centerline{\includegraphics[trim=0 0 0 0,width=2.0\columnwidth]{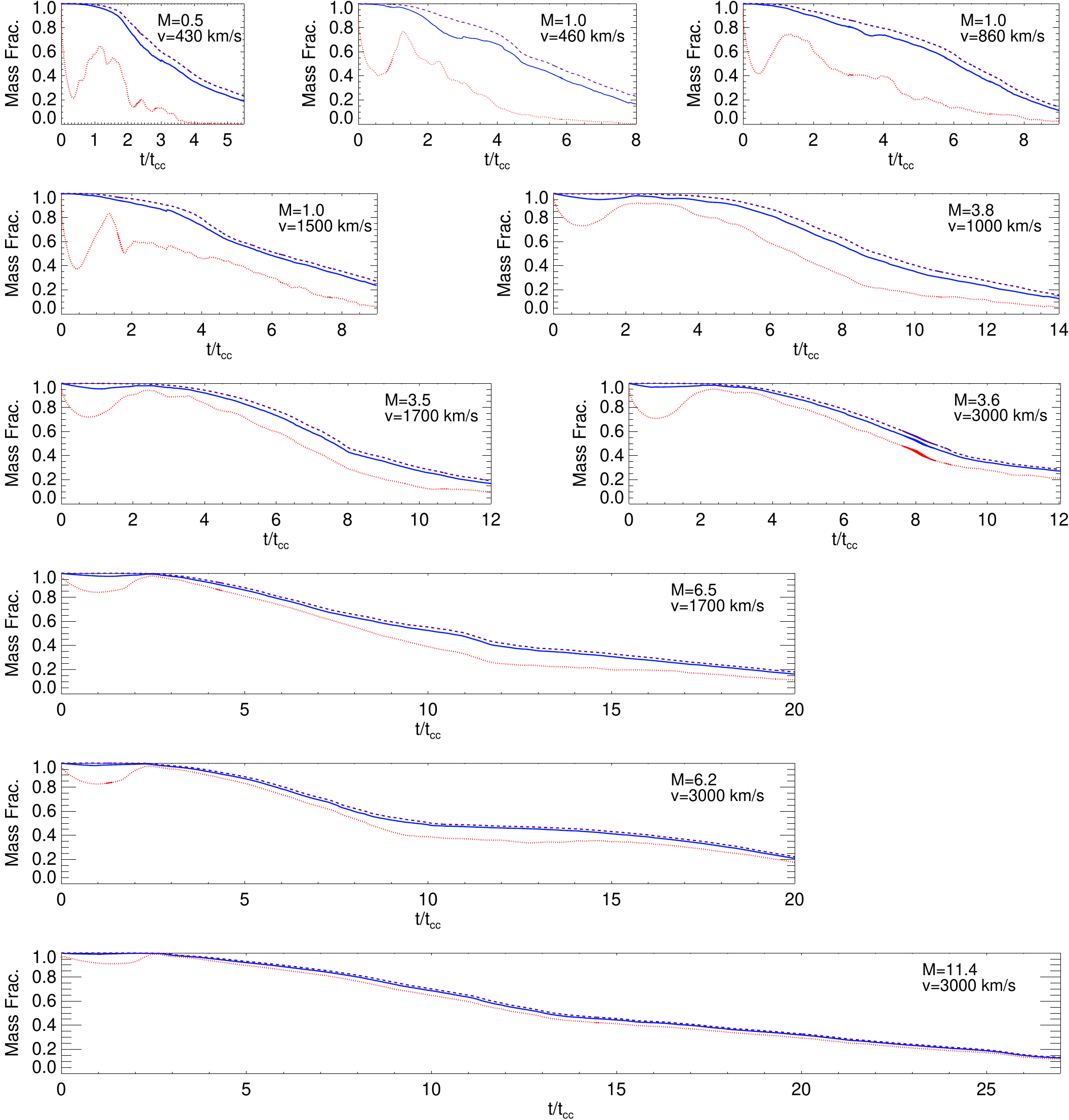}}
\caption{Mass evolution of the cloud as a function of mach number and velocity.   In each panel the solid blue line shows $F_{1/3},$ the fraction of the mass at or above 1/3 the original cloud density, the dotted red line shows $F_{1},$ the fraction mass at or above the original cloud density, and the purple dashed line shows $F_{1/10},$ the fraction of
the mass at or above 1/10 of the original density.   In all panels the time is given in units of the cloud crushing time, and the length of each panel is proportional to the time in $t_{\rm cc}$ units.\\}
\label{fig:massevolution}
\end{figure*}

\section{Results}
 
We carried out twelve simulations in all: ten runs with the same maximum resolution, which were used to span the parameter space in Table 1, and a two additional runs with different maximum resolutions, which were
used to examine resolution effects.   Each of our parameter study runs  was carried out on a base $64 \times 64 \times 48$ grid, with four additional levels of refinement.  The base resolution for these runs was 25 parsecs in each direction and the maximum resolution was $1.5625$ parsecs, or $R_{\rm cloud}/64.$  This value was chosen to be significantly smaller than the size advocated  by MacLow \& Zahnle (1994), who found that their 2D cylindrical simulations of the breakup of comet Shoemaker-Levy 9 in Jupiter's atmosphere required at least 25 zones across the projectile to achieve convergent results.  At this resolution, our simulations took 10-45k CPU hours to run, depending on the choice of parameters.

\subsection{Mass Evolution}

In Figure \ref{fig:massevolution}, we show the fraction of the mass retained by the cloud as a function of time in units of the cloud crushing time.   In this figure, the retained mass fraction is defined using three different measures: $F_1(t)$, the fraction of the total mass at or above the original cloud density, $F_{1/3}(t)$, the fraction of total mass at or above 1/3 the original cloud density, and $F_{1/10}(t)$, the fraction of the total mass at or above 1/10 of the original cloud density.  Although $F_1$  is the simplest measure of cloud mass, it displays a number  of misleading features.   Our simulations start with simple initial conditions, with the exterior medium moving at all points outside the cloud.  Thus the back of the cloud expands for roughly $0.5 t_{\rm cc},$ while the flow behind the cloud rearranges itself.   This causes all runs to display an initial drop in $F_1$ as a substantial part of the cloud drops slightly in density.    Conversely, from about $t_{\rm cc}$ to $2 t_{\rm cc},$ the shock at the front of the cloud increases the cloud density.   In the high Mach number runs, the density increase behind this (radiative) shock is large, bringing almost the entire cloud back to a density equal to or exceeding the original density, which brings  $F_1$ back to  $\approx 1$ by $t=2 t_{\rm cc}.$  At lower Mach numbers,  the density increase is somewhat smaller, such that the increase in $F_1$ is more modest.   

These transient features are avoided by working with $F_{1/3}$ or $F_{1/10}.$  In these cases, the initial downstream expansion and shock compression do not cause a significant change in the material interpreted as belonging to the cloud.   Instead, the cloud evolution is largely monotonic, remaining approximately constant for the initial stages of the interaction, and then dropping at a roughly constant rate at times $\gtrsim 2 t_{\rm cc}.$   This evolution is extremely similar between $F_{1/3}$ and $F_{1/10},$ indicating that the precise choice of density threshold does not affect our conclusions.  On the other hand, the timescale on which these mass fractions evolve varies extensively between the runs, such that  the cloud in the  $M_{\rm hot}=0.5,$ $v=430$ km/s  run has lost 3/4 of its mass by $t= 5 t_{\rm cc},$ but at this same time, the cloud in the $M_{\rm hot}=11.4,$ $v=3000$ km/s case still contains over 90\% of its original mass.

\begin{figure*}
\centerline{\includegraphics[trim=0 0 0 0,width=1.80\columnwidth]{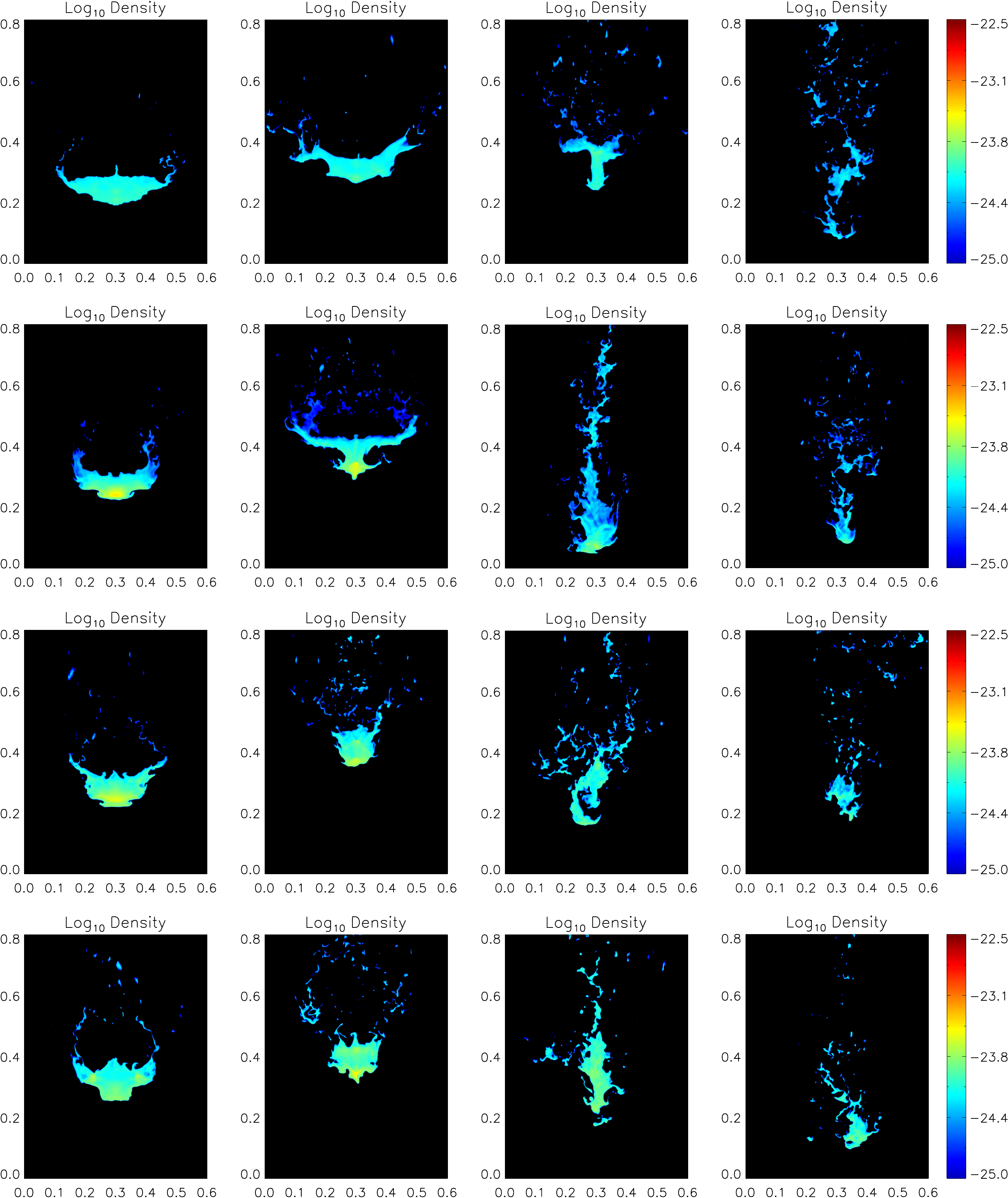}}
\caption{Plots of central density slices from simulations with Mach numbers 0.5 and 1, at times at which  the fraction of the mass at or above 1/3 the original density of the cloud is 90\% ($t_{90},$ first column), 75\% ($t_{75},$ second column), 50\% ($t_{50},$ third column), and 25\% ($t_{25},$ fourth column).  {\em First row:} Results from simulation M0.5v430,  at times  $t_{90}= 1.6 t_{\rm cc}$, $t_{75}= 2.2 t_{\rm cc}$, $t_{50} =  3.2 t_{\rm cc}$,  and $t_{25} =  4.8 t_{\rm cc}$. {\em Second row:} Results from simulation M1v480 at times  $t_{90} = 1.8 t_{\rm cc}$, $t_{75} =Ê2.6 t_{\rm cc}$, $t_{50} = 4.8 t_{\rm cc}$, and $t_{25} =  7.0 t_{\rm cc}.$ {\em Third row:} Results from simulation M1v860 at times       $t_{90} = 2.0 t_{\rm cc}$, $t_{75} =       3.4 t_{\rm cc}$, $t_{50} =       6.0 t_{\rm cc}$, and $t_{25} =  7.6 t_{\rm cc}.$  {\em Fourth row:} Results from simulation M1v1500 at times $t_{90} = 2.4 t_{\rm cc}$, $t_{75} =       3.8 t_{\rm cc}$, $t_{50} =  5.8 t_{\rm cc}$, and $t_{25} =   8.8 t_{\rm cc}$. All densities are in g cm$^{-3}$ and all lengths are in kpc.  These figures, along with the slice plots below, are insets of the region 0.6 $\times$ 0.8 kpc region around the cloud, which is a subset of the full $1.6 \times 1.6 \times1.2$ kpc simulation volume.}
\label{fig:M0.5and1}
\end{figure*}

\begin{figure*}
\centerline{\includegraphics[trim=0 0 0 0,width=2.0\columnwidth]{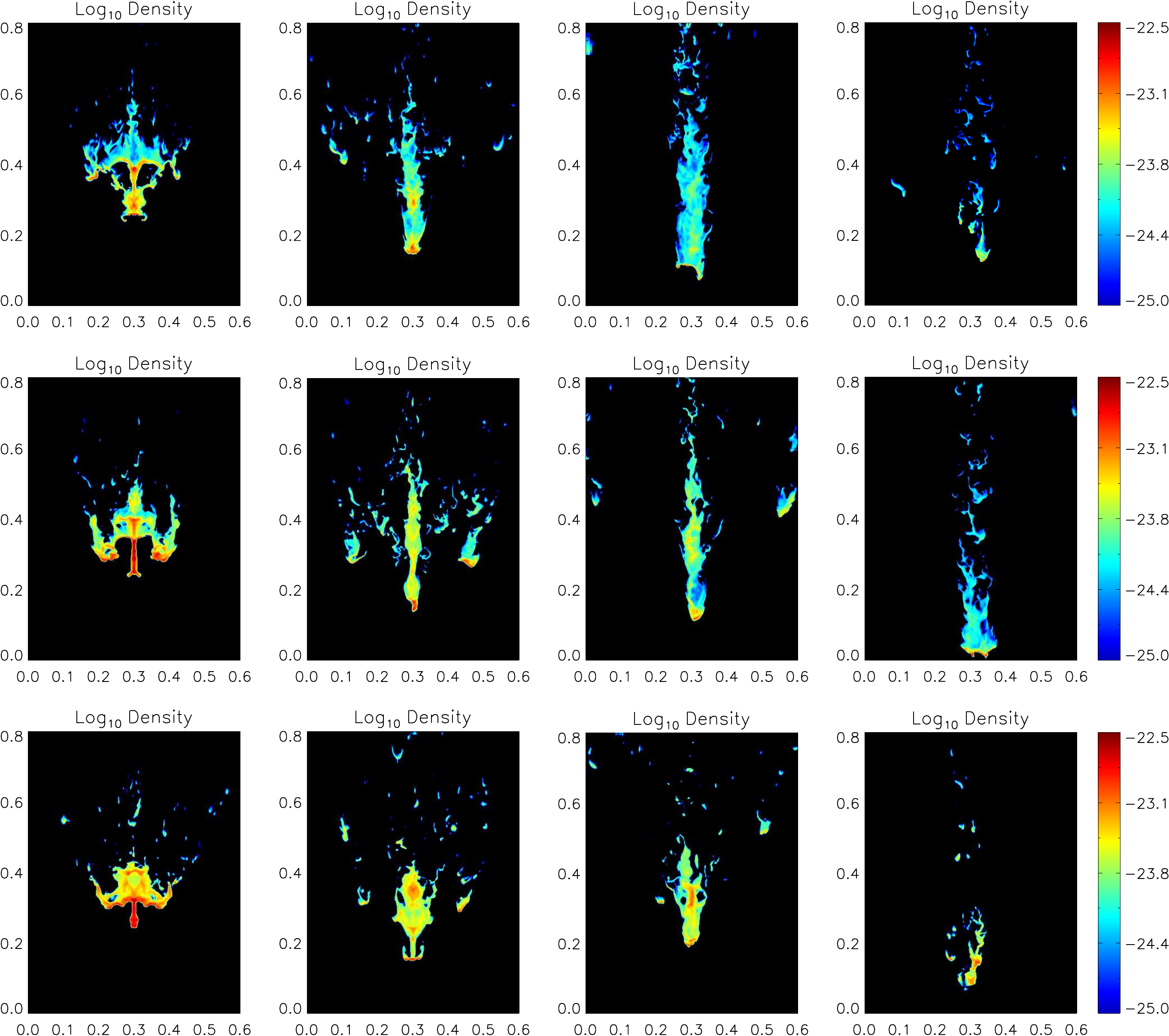}}
\caption{Plots of central density slices from simulations with Mach numbers $\approx 3.6.$  As in Figure \ref{fig:M0.5and1}, the results are shown at times at which  the fraction of the mass at or above 1/3 the original density of the cloud is 90\% ($t_{90},$ first column), 75\% ($t_{75},$ second column), 50\% ($t_{50},$ third column), and 25\% ($t_{25},$ fourth column).  {\em First row:} Results from simulation M3.8v1000 at times  $t_{90} = 5.0 t_{\rm cc}$, $t_{75} =      6.6 t_{\rm cc}$, $t_{50} =  8.4 t_{\rm cc}$, and $t_{25} = 11.6 t_{\rm cc}.$ {\em Second row:} Results from simulation M3.5v1700 at times  $t_{90} = 4.2 t_{\rm cc}$, $t_{75} =      5.8 t_{\rm cc}$, $t_{50} =  7.6 t_{\rm cc}$,  and  $t_{25} =  10.4 t_{\rm cc}.$ {\em Third row:} Results from simulation M3.6v3000 at times $t_{90} = 4.2 t_{\rm cc}$, $t_{75} =      6.0 t_{\rm cc}$, $t_{50} =  8.0 t_{\rm cc}$,  and $t_{25} =  12.0 t_{\rm cc}.$ All densities are in g cm$^{-3}$ and all  lengths are in kpc.}
\label{fig:M3.5}
\end{figure*}

\begin{figure*}
\centerline{\includegraphics[trim=0 0 0 0,width=2.0\columnwidth]{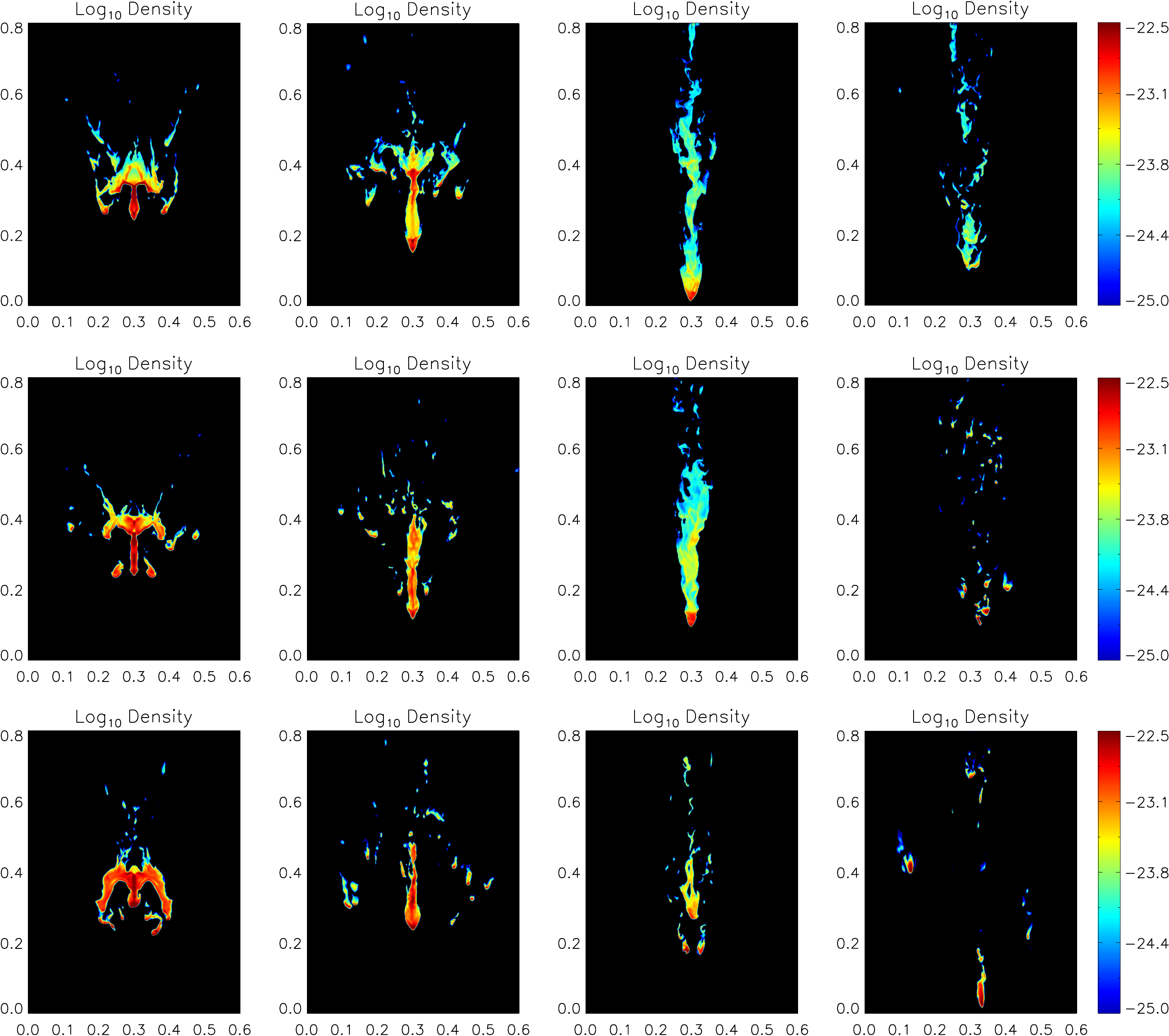}}
\caption{Plots of central density slices from simulations with Mach numbers $\approx 6.5$ and 11, with times as in Figures \ref{fig:M0.5and1} and \ref{fig:M3.5} {\em First row:} Results from simulation M6.5v1700 at times $t_{90} = 4.4 t_{\rm cc},$ $t_{75} =   6.4 t_{\rm cc},$ $t_{50} =   10.6 t_{\rm cc},$ and $t_{25} =  17.0 t_{\rm cc}.$ {\em Second row:} Results from simulation M6.2v3000 at times $t_{90} = 4.4 t_{\rm cc},$ $t_{75} =   6.4 t_{\rm cc},$ $t_{50} =   9.6 t_{\rm cc},$  and $t_{25} =  19.0 t_{\rm cc}.$ {\em Third row:} Results from simulation M11.4v3000 at times $t_{90} = 5.6 t_{\rm cc},$ $t_{75} =   8.8 t_{\rm cc},$ $t_{50} =   13.0 t_{\rm cc},$ and $t_{25} =  22.2 t_{\rm cc}.$ All densities are in g cm$^{-3}$ and all  lengths are in kpc.\\}
\label{fig:M6.5_11}
\end{figure*}

To better illustrate the evolution of the clouds, in Figures \ref{fig:M0.5and1}, \ref{fig:M3.5},  and \ref{fig:M6.5_11} we plot slices of the central density distribution at four characteristics times: $t_{\rm 90},$ the time at which $F_{1/3} = 90\%$, $t_{\rm 75},$ the time at which  $F_{1/3} = 75\%,$ $t_{\rm 50},$ the time at which $F_{1/3} = 50\%,$ and $t_{\rm 25},$ the time at which $F_{1/3} = 25\%.$    In the lowest Mach number runs, illustrated in  Figure \ref{fig:M0.5and1}, $t_{90}$ usually occurs during the early stages of the interaction, when $t \approx 2 t_{\rm cc}$.  For example, in the $M_{\rm hot}=0.5,$ $v=430$ km/s run, the cloud loses 10\% of its mass by $t=1.6 \, t_{\rm cc},$ mostly due to material stripped from the sides during the initial interaction.   As time goes on, this run evolves primarily through the growth of the Kelvin-Helmholtz instability, which works its way in from the sides at a rate $\propto v_{\rm hot}/\chi_0^{1/2}$, disrupting the majority of the cloud by $5 t_{\rm cc}.$

As expected from the discussion in \S2.2, the growth of the KH instability is somewhat slowed in the  $M_{\rm hot}=1$ case,  such that $ t_{25} \approx 8 t_{\rm cc}$ with slightly higher values being seen in the highest velocity cases.   This evolution is extended even further in the Mach $\approx$ 3.5 runs,  which retain over 90\% of their mass until well after the initial stages of the interaction.  Thus the slices taken at $t_{\rm 90}$ in Figure \ref{fig:M3.5} show the clouds in highly shocked and compressed configurations, with numerous features that have densities well above the initial density of $\rho = 10^{-24}$ g cm$^{-3}$.  As time evolves in the supersonic runs,  these clouds become more cometary in appearance, with the densest features located near the head of the cloud, which is trailed by a long tail of lower density material.   This tail is then slowly mixed into the surrounding medium such that most of the mass is lost by  $10 t_{\rm cc}$.   

These features are exaggerated even further in the Mach $\approx 6.5$ and 11 cases, shown in Figure \ref{fig:M6.5_11}.   Here the $t_{\rm 90}$ plots, taken at $\approx 5 t_{\rm cc},$ show regions with densities exceeding 30 times the original density.   Again these clouds evolve into cometary distributions with moderate density contrasts, and these retain a large fraction of their mass for even longer than the $M_{\rm hot} \approx3.5$ runs, such that $t_{25} \approx 20 t_{\rm cc}.$  Note that these long disruption times occur at $t/t_{\rm cc}$ values later than have been simulated previously.  On the other hand, they occur {\em sooner} than expected from the laboratory measurements, especially as fit by Slessor \etal (2000) with eq.\ (\ref{eq:Slessor}).

\begin{figure}
\centerline{\includegraphics[trim=0 0 0 0,width=1.0\columnwidth]{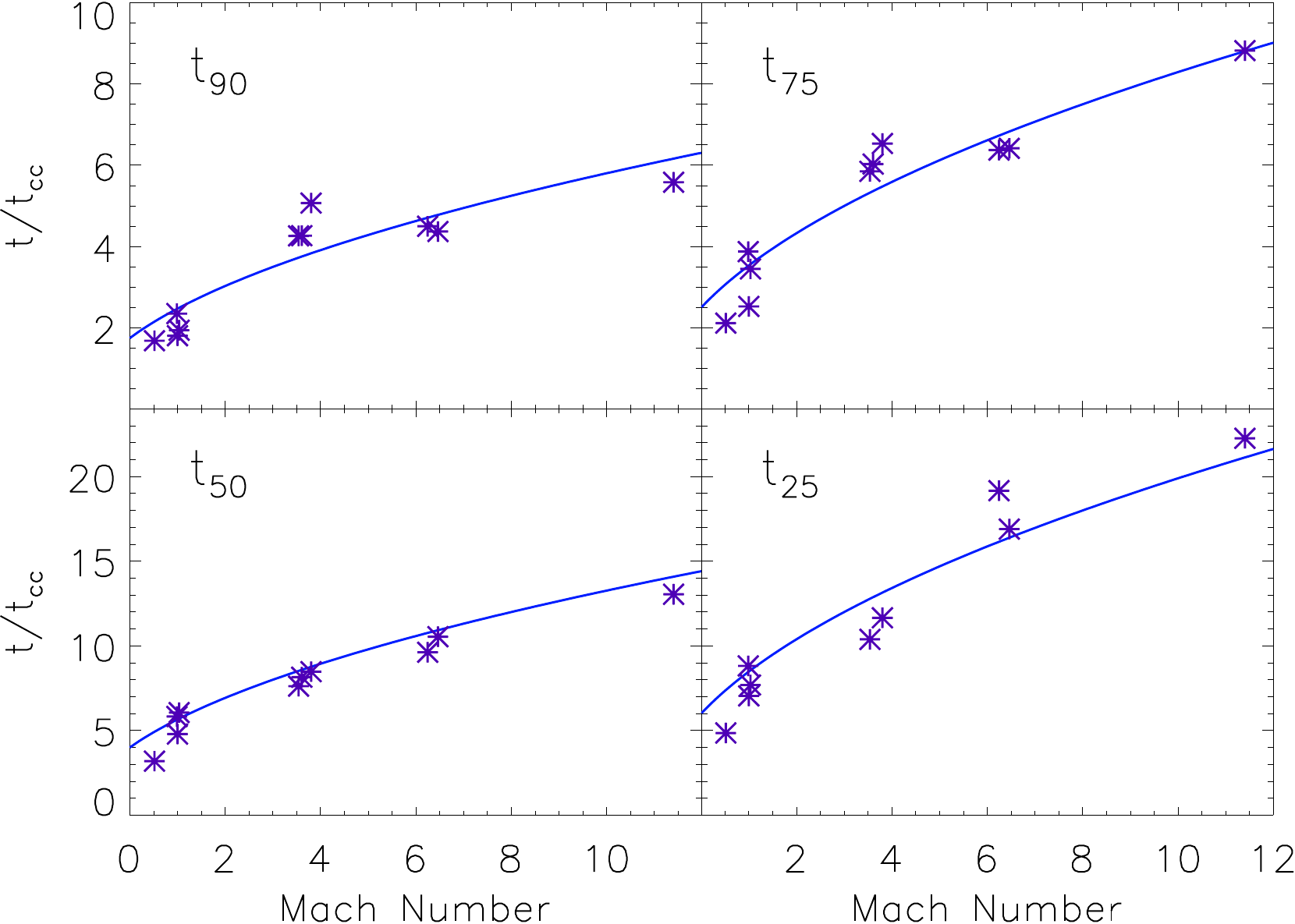}}
\caption{Time at which the fraction of the mass at or above 1/3 the original density of the cloud is  90\% ($t_{90}$), 75\% ($t_{75}$),  50\% ($t_{50}$), 25\% ($t_{25}$) in units of the cloud crushing time. The (violet) points show the simulation results, and the (blue) lines show the fits $t_{90} \approx 1.75 t_{\rm cc}  \sqrt{1+M_{\rm hot}},$ $t_{75} \approx 2.5 t_{\rm cc} \sqrt{1+M_{\rm hot}},$ $t_{50} \approx 4 t_{\rm cc} \sqrt{1+M_{\rm hot}},$ and $t_{25} \approx 6 t_{\rm cc}\sqrt{1+M_{\rm hot}},$   respectively.}
\label{fig:timescales}
\end{figure}

In Figure \ref{fig:timescales}, we show $t_{90}$, $t_{75}$,  $t_{50}$, and $t_{25}$ in units of cloud crushing times for each of our runs, plotted as a function of Mach number. In all cases, we find that the scatter between runs with difference $\chi_0$ values but the same Mach number is small, meaning that the differences in disruption times at constant $M_{\rm hot}$ is well accounted for  by the $\chi_0^{1/2}$ scaling of $t_{\rm cc}.$   Similarly, the scaling with Mach number is well fit by a simple function, such that in all panels we obtain good agreement with our simulations results and timescales $\propto \sqrt{1+M_{\rm hot}}.$ In particular, we find that our results match  
\be
t= \alpha t_{\rm cc} \sqrt{1+M_{\rm hot}}
\label{eq:tfit}
\ee
where $\alpha $= 1.75, 2.5, 4, and 6, at $t_{90},$ $t_{75},$ $t_{50},$ and $t_{25},$ respectively.

\begin{figure*}
\centerline{\includegraphics[trim=0 0 0 0,width=2.0\columnwidth]{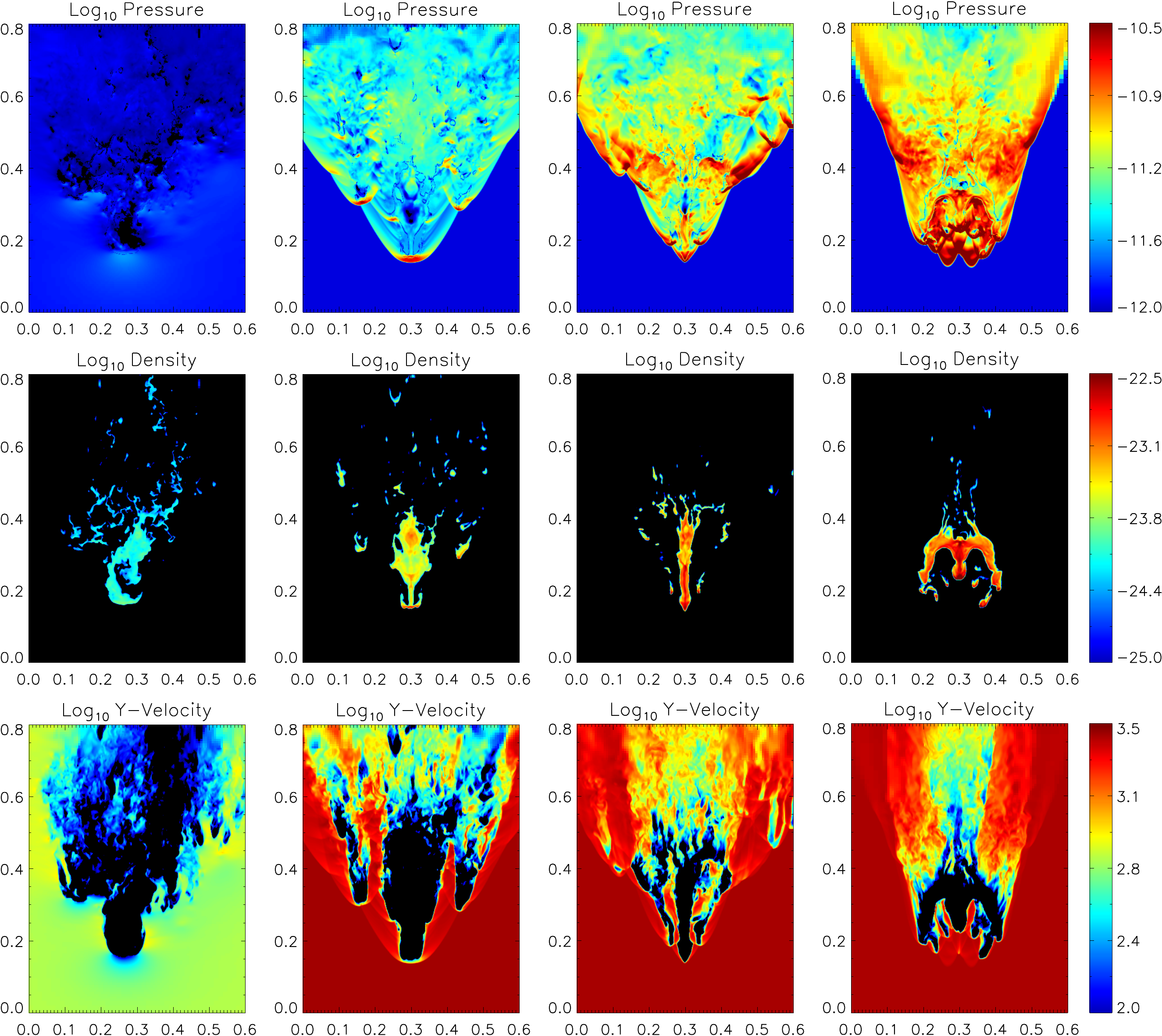}}
\caption{Slices of pressure in units of ergs cm$^{-3}$ (top row), density in units of g cm$^{-3}$ (center row), and velocity in units of km s$^{-1}$ (bottom row), taken at the midplane at six cloud crushing times.  The slices are taken from runs M1v860 (first column), M3.6v3000 (second column), M6.2v3000 (third column), and M11.4v3000 (fourth column). All lengths are in kpc.\\}
\label{fig:pdv.pdf}
\end{figure*}

To better understand the origin of this scaling, in Figure \ref{fig:pdv.pdf}, we plot slices of the pressure, density, and velocity distribution from runs with four different Mach numbers:  an $M_{\rm hot}=1$ run with $v_{\rm hot} = 840$ km/s, and three runs with $v_{\rm hot} = 3000$ km/s and $M_{\rm hot}=3.6,$ 6.2, and 11.4.  For ease of comparison between runs, all slices in this figure are plotted at the same point in their evolution, when $t/t_{\rm cc} = 6.$  From the top panels in this figure, we can directly compare the pressure distribution established in each of these runs.   Here we see that the increase in pressure is strongest at the front of the cloud, with values  approaching the $\approx 1+M_{\rm hot}^2$ increase expected for a normal shock [see eq.\ (\ref{eq:p31})]. 

However, downstream from the front of the cloud the pressure increase is more modest, scaling instead as $\approx 1 + M.$ This weaker scaling with Mach number can be understood by the fact that the majority of the cloud material does not encounter flow material that has passed through a normal shock, but rather through an oblique shock.   In the case in which flow is passing over a wedge with opening angle $\theta,$ an oblique shock is formed with a  downstream pressure given by
\be
\frac{p_2}{p_1} = 1+\frac{\gamma M^2 \theta}{\sqrt{M^2-1}} \approx 1 + \gamma \theta M
\ee
(Ackert 1925), rather than $\approx M^2$.   While this is a somewhat idealized example, the strong alignment between the shock front and the direction of the hot material suggests that  the flow is much better approximated by an oblique shock encountering a $\theta \approx 1/\gamma \approx 30^\circ$  wedge than a normal shock encountering
a $ \theta  \approx 90^\circ$ wall.  This large difference in pressure between oblique and normal shocks also highlights the importance of adopting a large simulation domain, such that the zero normal gradient assumed for all flow variable at the boundaries does not affect the shape of the shock.

By comparing the pressure slices to the density slices, we can also see that the downstream shock is well-separated from the edge of the cloud.   The primary role of the shock throughout the majority of the cloud is to increase its density by $\approx 1 +  M$, as this material must  rearrange itself to achieve pressure equilibrium with the exterior flow while still maintaining a constant temperature.   Thus a large pressure gradient develops between the head and tail of the cloud, which leads to an accelerating expansion of the cloud in the $z$ direction by a rate
\be
\Delta v_z(t)  =  \frac{t}{\rho_{\rm cloud}}  \frac{d P}{dR} \approx  t \frac{\rho_{\rm hot} v_{\rm hot}^2 }{\rho_{\rm cloud} R_{\rm cloud} } = \frac{t}{t_{\rm cc}} \frac{v_{\rm hot}}{(1+M_{\rm hot}) \chi_0^{1/2}},
\ee
such that 
\be
\frac{R_z(t)}{R_{\rm cloud}}  \approx   \frac{1}{2} \left( \frac{ t}{t_{\rm cc} \sqrt{1+M_{\rm hot}}} \right)^2.
\ee
This stretching in the direction of the flow is the source of the cloud's cometary appearance at late times in the high $M_{\rm hot}$ runs.
Furthermore, from our empirical fit in eq.\ (\ref{eq:tfit}) we find that  $R_z \approx$  is almost purely a function of the
mass lost from the cloud, such that $R_z \approx$  $1.5 R_{\rm cloud},$ $3 R_{\rm cloud},$ $8R_{\rm cloud},$
and $18 R_{\rm cloud}$ at $t_{90},$ $t_{75},$ $t_{50},$ and $t_{25},$ respectively.

These estimates are roughly consistent with the lengths seen in the density slices, with the exception of the distributions at $t_{25},$ in which it is very difficult to assign lengths to the clouds, which have become fragmented into many small clumps. Thus, even as the KH instability is strongly damped in these supersonic runs, the clouds are also subjected to rapid stretching by the streamwise pressure gradient.  This first distorts the clouds, then finally splits them apart as $R_z(t)/{R_{\rm cloud}}$  grows rapidly  on a timescale $\propto t_{\rm cc} \sqrt{1+M_{\rm hot}}.$

\subsection{Velocity Evolution}

\begin{figure*}
\centerline{\includegraphics[trim=0 0 0 0,width=2.0\columnwidth]{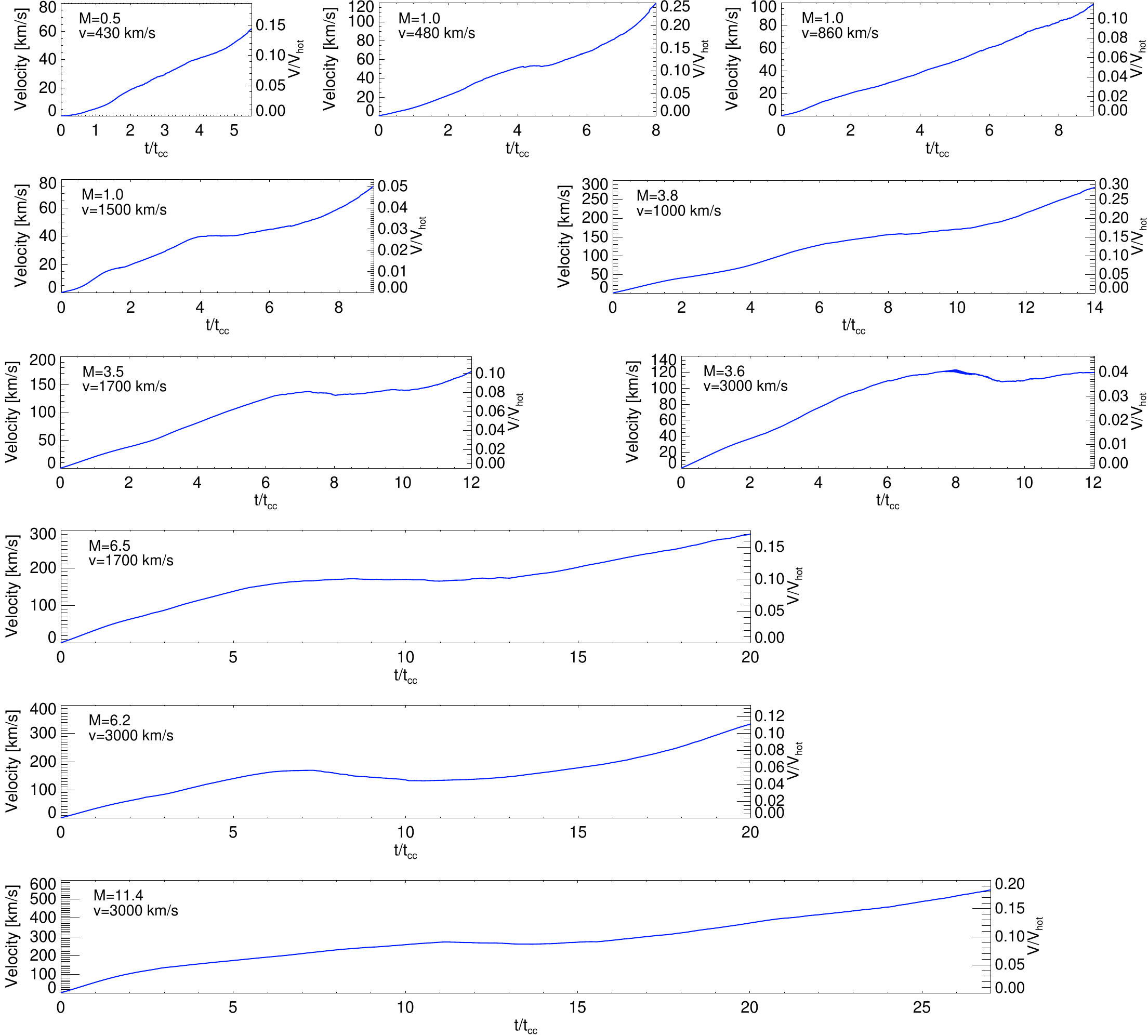}}
\caption{Velocity evolution of the cloud as a function of Mach number and velocity.   In each panel the solid (blue) line shows the mass weighted average velocity of the mass at or above 1/3 the original density of the cloud As in figure \ref{fig:massevolution}, in all panels the time is given in units of the cloud crushing time, and the length of each panel is also proportional to $t/t_{\rm cc}$.\\}
\label{fig:velocity}
\end{figure*}

Next we consider the velocity evolution of the cloud, comparing our simulation results to the simple scalings expected from eq.\ (\ref{eq:vcoft}) above. Because the primary impact of the shocks seen at high $M_{\rm hot}$ values is to compress the cloud, this means that the extent of the cloud perpendicular to the direction of the shock,  $\tilde R_{\perp}(t) \equiv R_{\perp}(t)/R_{\rm cloud},$ will drop as the cloud is squeezed by the pressure increase behind it.    Thus, although the high Mach number cases are able to avoid disruption for many more cloud crushing times, the acceleration of such clouds per cloud crushing time will also be smaller. In Figure \ref{fig:velocity}, we plot the velocity evolution of each our simulations, using units of both km/s and $v_{\rm hot}$.    Here we see that, while the runs with the highest Mach numbers tend to be those with the highest overall final velocities, there are not necessarily the runs with the highest value of $v_{\rm cloud}/v_{\rm hot}$, meaning that the decrease of  $\tilde R_{\perp}$ often has a bigger effect on the final velocity of the cloud than the increase in the overall cloud lifetime.

\begin{figure}
\centerline{\includegraphics[trim=0 0 0 0,width=1.0\columnwidth]{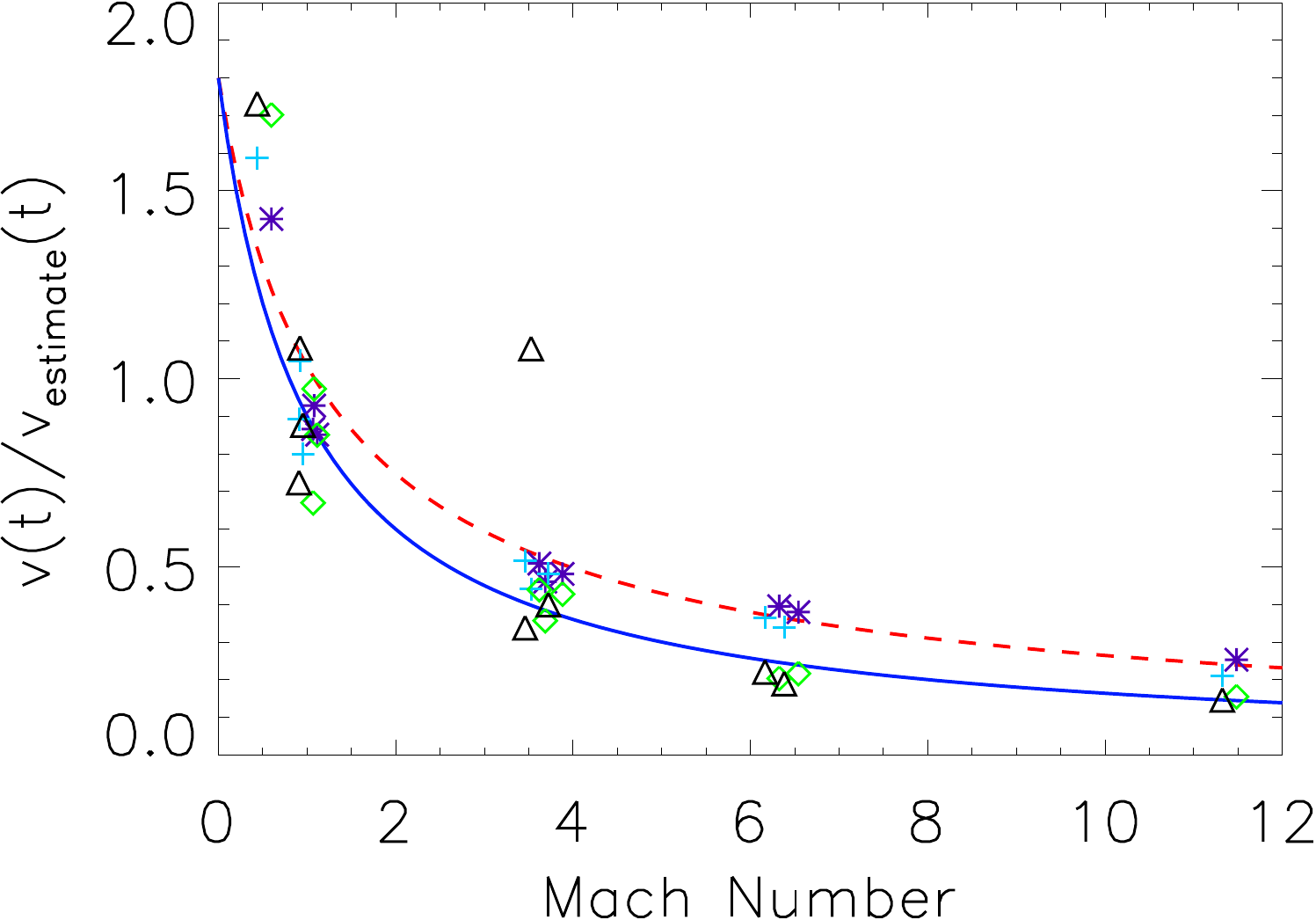}}
\caption{Velocity evolution of the cloud normalized by $v_{\rm estimate}(t) \equiv \frac{3 v_{\rm hot} t}{4 \chi_0^{1/2} t_{\rm cc}}$ the expected value if  the velocity of the exterior medium relative to the cloud, the radius of the cloud, and the density contrast between the cloud and the exterior medium all remain fixed.  The points are velocities at $t_{90}$ (violet asterisks), $t_{75}$ (cyan pluses), $t_{50}$ (green diamonds),  $t_{25}$ (black triangles). The blue solid curve shows $1.8 (1+M_{\rm hot})^{-1}$ and the red dashed curve shows $1.8 (1+M_{\rm hot})^{-0.8}.$  \\ }
\label{fig:velocityevolution}
\end{figure}

To study this  evolution in greater detail, we normalized the cloud velocity in each case by the expectations from eq.\ (\ref{eq:vcoft}) in the absence of changes in 
$\Delta \tilde v$ and $\tilde R_{\perp}$, computing $v_{\rm cloud}(t)/v_{\rm estimate}(t)$ with
\be
v_{\rm estimate}(t)  \equiv \frac{3 v_{\rm hot} t}{4 \chi_0^{1/2} t_{\rm cc}}. 
\ee
These ratios are plotted as a function of Mach number in Figure \ref{fig:velocityevolution} at each of the characteristic times defined above: $t_{90},$ $t_{75}$,  $t_{50}$,  and $t_{25},$ each plotted with a different symbol.     This comparison shows that while $v_{\rm cloud}/v_{\rm estimate}$ varies significantly between runs, these differences,  like the timescales in \S4.1, are functions almost exclusively of $M_{\rm hot}.$   Furthermore, the dependence of $v_{\rm cloud}/v_{\rm estimate}$ is also only weakly dependent on the time at which the velocities are measured: with the earlier times being well described by $v_{\rm cloud}/v_{\rm estimate} = 1.8 (1+M_{\rm hot})^{-0.8}$ and the late times being well described by  $v_{\rm cloud}/v_{\rm estimate} =1.8 (1+M_{\rm hot})^{-1}.$  

These scalings can be understood in terms of changes in  $\tilde R_{\perp},$ which means that a density increase by a factor $\approx 1+M_{\rm hot}$ will have a  different effect on the cloud evolution depending on the direction of the contraction.  If the contraction is purely in the direction of the flow, then  $\tilde R_\perp$  will remain unchanged, and the contraction will not affect $v_{\rm cloud}/v_{\rm estimate}.$  On the other hand, a uniform contraction in density by a factor of $1+M$ will result in $\tilde R_\perp \propto M^{-1/3}$  such that  $v_{\rm cloud}/v_{\rm estimate} \propto (1+M)^{-2/3}.$  Finally, a contraction purely perpendicular to the flow will result in $v_{\rm cloud}/v_{\rm estimate} \propto (1+M_{\rm hot})^{-1}.$ 

Both the $v_{\rm cloud}/v_{\rm estimate} =1.8 (1+M_{\rm hot})^{-0.8}$ and the $v_{\rm cloud}/v_{\rm estimate} =1.8 (1+M)^{-1}$ models plotted in Figure \ref{fig:velocityevolution}   exceed 1 in the case in which $M_{\rm hot} \leq 1.$  This means that they correspond to a {\em increase} in $\tilde R_\perp.$   In fact, a moderate lateral expansion, due to the initial shock that passes through the cloud, is visible in the low Mach number slice plots shown in Figure \ref{fig:M0.5and1}.  Comparing these slices with the higher  $M_{\rm hot}$ runs in Figures \ref{fig:M3.5} and \ref{fig:M6.5_11}, we can also see that the compression of the cloud at higher Mach number is somewhat biased to the direction to the flow at early times, resulting in a $v_{\rm cloud}/v_{\rm estimate}$ scaling between $(1+M_{\rm hot})^{-2/3}$ and $(1+M_{\rm hot})^{-1}.$   At late times, the expansion of the cloud in the streamwise direction means that the compression is almost
completely perpendicular to the flow direction, resulting in a scaling that goes as $\approx (1+M_{\rm hot})^{-1}.$

\begin{figure}
\centerline{\includegraphics[trim=0 0 0 0,width=1.0\columnwidth]{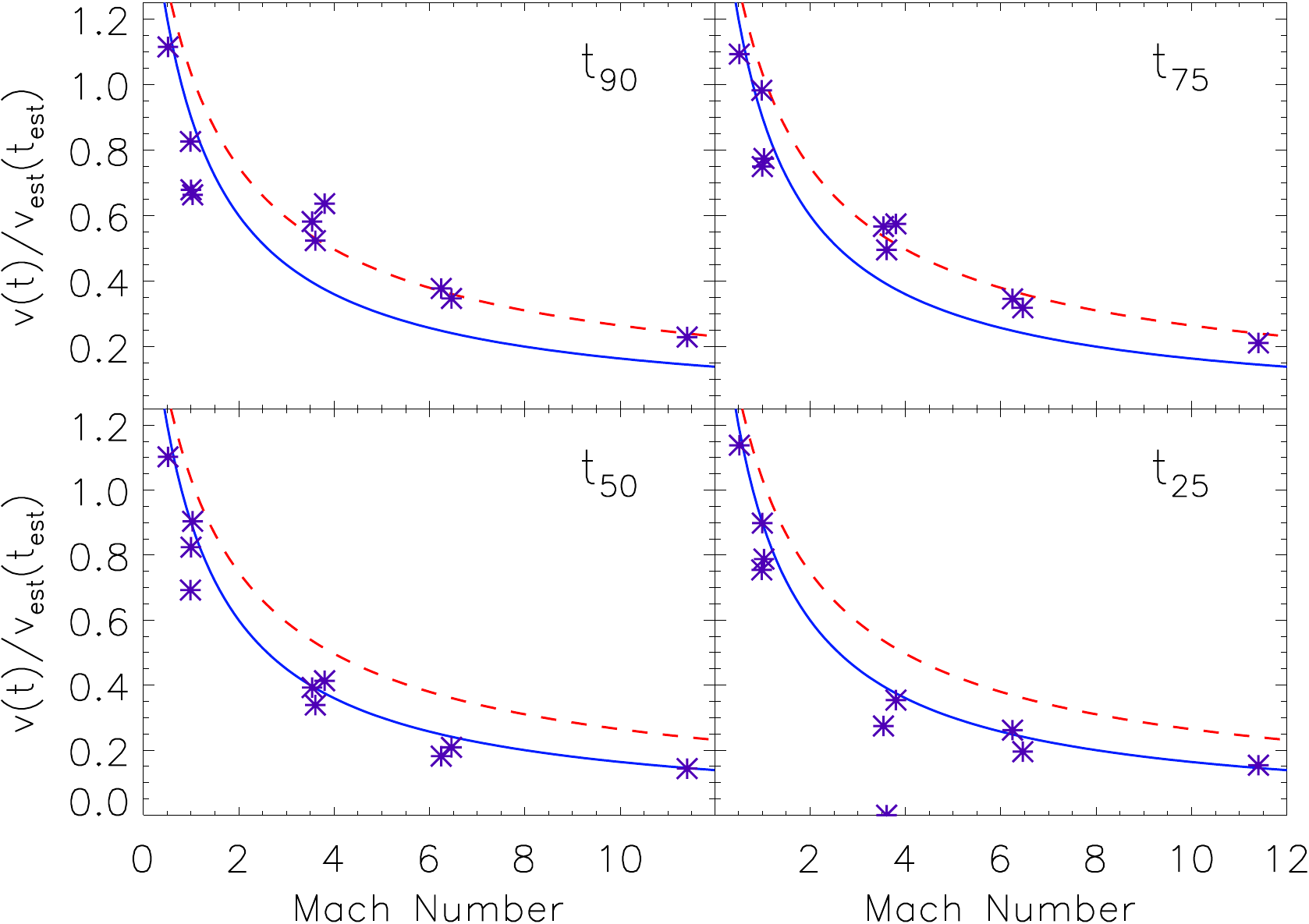}}
\caption{Velocity evolution of the cloud at times $t_{90}$ (top left), $t_{75}$ (top right),  $t_{50}$ (bottom left), $t_{25}$ (bottom right), normalized by  $v_{\rm estimate}(t) \equiv \frac{3 v_{\rm hot} t}{4 \chi_0^{1/2} t_{\rm cc}}$  evaluated at the {\em estimated} values of these times as given by equation (\ref{eq:tfit}).  As in Figure \ref{fig:velocityevolution}, the violet points are the simulation results,  the blue solid curve shows $1.8 (1+M_{\rm hot})^{-1}$ and the red dashed curve shows $1.8 (1+M_{\rm hot})^{-0.8}.$    The more shallow curve provides the best fit at the earliest times, but at late times the results are better fit with the stronger scaling with Mach number.\\}
\label{fig:velocityevolution2}
\end{figure}

In Figure \ref{fig:velocityevolution2} we combine our fits for the mass loss and $v_{\rm cloud}(t)$ to obtain
\be
v_{\rm cloud}(t) = \frac{3 v_{\rm hot} [\alpha (1+M_{\rm hot})^{1/2}] }{4 \chi_0^{1/2}} \frac{1.8}{(1+M_{\rm hot})^\beta} 
\label{eq:vcloud}
\ee
where as in eq.\  $\alpha$ = 1.75, 2.5, 4, and 6, at $t_{90},$ $t_{75},$ $t_{50},$ and $t_{25},$ respectively and $\beta = 0.8$ or $1.$  This expression provides a good fit to all our data,
with $\beta=0.8$  fitting better at $t_{90}$ and $t_{75}$ and $\beta=1$ fitting better at late times.  Together, eqs.\ (\ref{eq:tfit}) and (\ref{eq:vcloud}) provide a good summary of our full set of results for the evolution of clouds in the hydrodynamic case in which thermal conduction is small.

\subsection{Distances and Implications for Circumgalactic Obsevations}

\begin{figure}
\centerline{\includegraphics[trim=0 0 0 0,width=1.0\columnwidth]{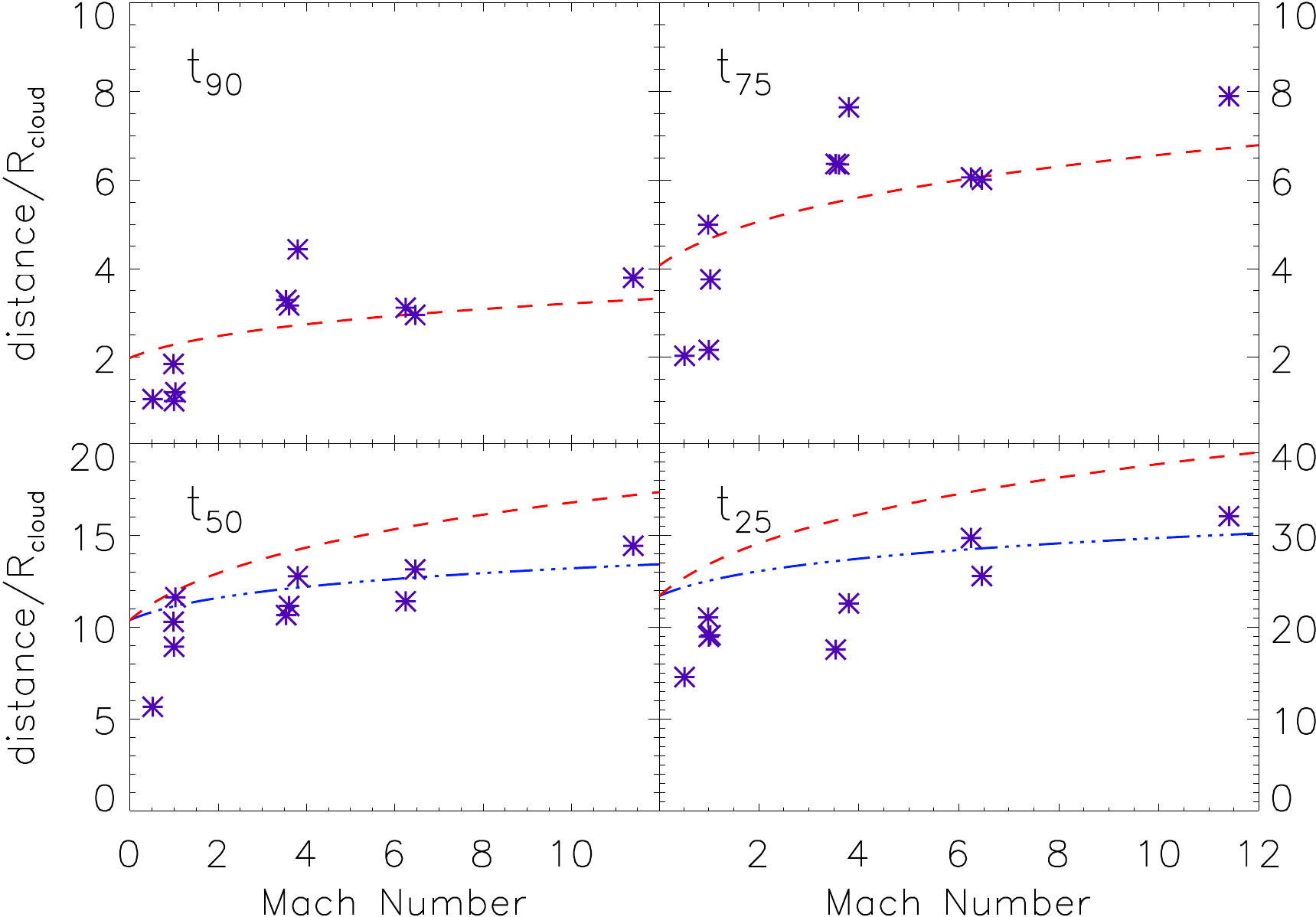}}
\caption{Distance traveled by the clouds as a function of the disrupted mass.  The distances scale with cloud radius and thus are expressed in units $R_{\rm cloud}$, which was 100 parsecs in our current simulations.  The data is compared to eq.\ (\ref{eq:d}) with $\beta=0.8$ (red dashed curves) and $\beta=0.9$ (blue dot-dashed curves), with $\alpha $= 1.75, 2.5, 4, and 6.}
\label{fig:distances}
\end{figure}

Although conduction and magnetic fields will both play important roles in full models of cold cloud evolution, we can nevertheless make preliminary comparisons between our results and current observations.  From eq.\ (\ref{eq:vcloud}),  we can see that the cloud velocity goes as $v_{\rm cloud} \approx 1.4 \alpha v_{\rm hot} \chi_0^{-1/2} (1+M)^{0.5-\beta},$ with $\alpha=6$ when the clouds are down to 25\% of their original mass, rapidly shedding their remaining gas.   The immediate implication is that if $\chi_0 \gtrsim 100$ {\em no clouds will ever be accelerated to the hot wind speed before disruption.}  Furthermore, typical values of $v_{\rm cloud}$ reach maximum values of $0.3 v_{\rm hot}$ or substantially less.     

Eq.\ (\ref{eq:vcloud}), also shows that the cloud velocity at a given stage in its evolution depends almost completely on the velocity of the hot wind and the initial density contrast between the cloud and the exterior medium, with the largest velocities  arising when $v_{\rm hot}$ is large and $\chi_0$ is small.   To reach the $\gtrsim 200$ km/s outflow velocities seen around large, rapidly star forming galaxies, (\eg Heckman \etal 2000; Pettini 2001; Rupke \etal 2002; Martin 2005, 2012) thus requires that the clouds enter the hot medium outside of the driving region, where the wind velocities are the largest and the pre-shock ambient pressures are the smallest.   

Integrating  our fit to $v_{\rm cloud}$ out to the disruption times measured above, we can also predict the distance that the cloud will travel at different points in its evolution.
This gives
\be
d_{\rm cloud} =  0.65 \alpha^2 R_{\rm cloud} (1+M)^{(1-\beta)},
\label{eq:d}
\ee
with $\alpha$ and $\beta$ defined as above.      In Figure \ref{fig:distances} we compare this estimate with our simulation results.  While the simulated distances show a bit more scatter than our velocity and timescale measurements, eq.\ (\ref{eq:d}) nevertheless gives a reasonable fit to the data, with $\beta=0.8$ providing a good description at the earliest times, and $\beta=0.9,$ which averages over the velocity evolution seen at early and late times, providing a better fit at late times.    As both of these $\beta$ values are very near one, the distance the cloud travels as a function of mass loss remans roughly constant over a large range of external conditions.

This means that, at least in the hydrodynamic case with no conduction, the distance traveled by the clouds depend almost exclusively only the initial cloud radius, with the clouds becoming completely disrupted as they move beyond  $\approx 40 R_{\rm cloud}.$    These distances are large enough that, given initial radii $\approx 100$ parsecs, cold clouds can travel from near the edge of the driving region to the few kpc distances at which they are typically observed in absorption against the starbursting host galaxy  (\eg Heckman \etal 2000; Pettini 2001; Soto \& Martin 2012).  

On the other hand, it is much more difficult for clouds to travel $\approx 100$ kpc distances, as probed in nearby galaxies using absorption lines measured in background quasars and galaxies (Bergeron 1986; Lanzetta \& Bowen 1992; Steidel \etal 1994, 2002, 2010;  Zibetti \etal 2007; Kacprzak \etal 2008; Chen \etal 2010; Tumlison \etal 2013; Werk \etal 2013, 2014; Peeples \etal 2014; Turner \etal 2014).   Our results show that, regardless of the structure of the wind, such cold clouds would have to be the sizes of entire galaxies to travel to such large distances without being disrupted.   Thus, unless conduction and magnetic effects are able to preserve clouds for much longer than seen our current simulations, the $\approx 10^4$K gas observed at $\approx 100$ kpc around galaxies can not be directly associated with $10^4$K ejected material.

\subsection{Resolution Effects}

\begin{figure*}
\centerline{\includegraphics[trim=0 0 0 0,width=2.05\columnwidth]{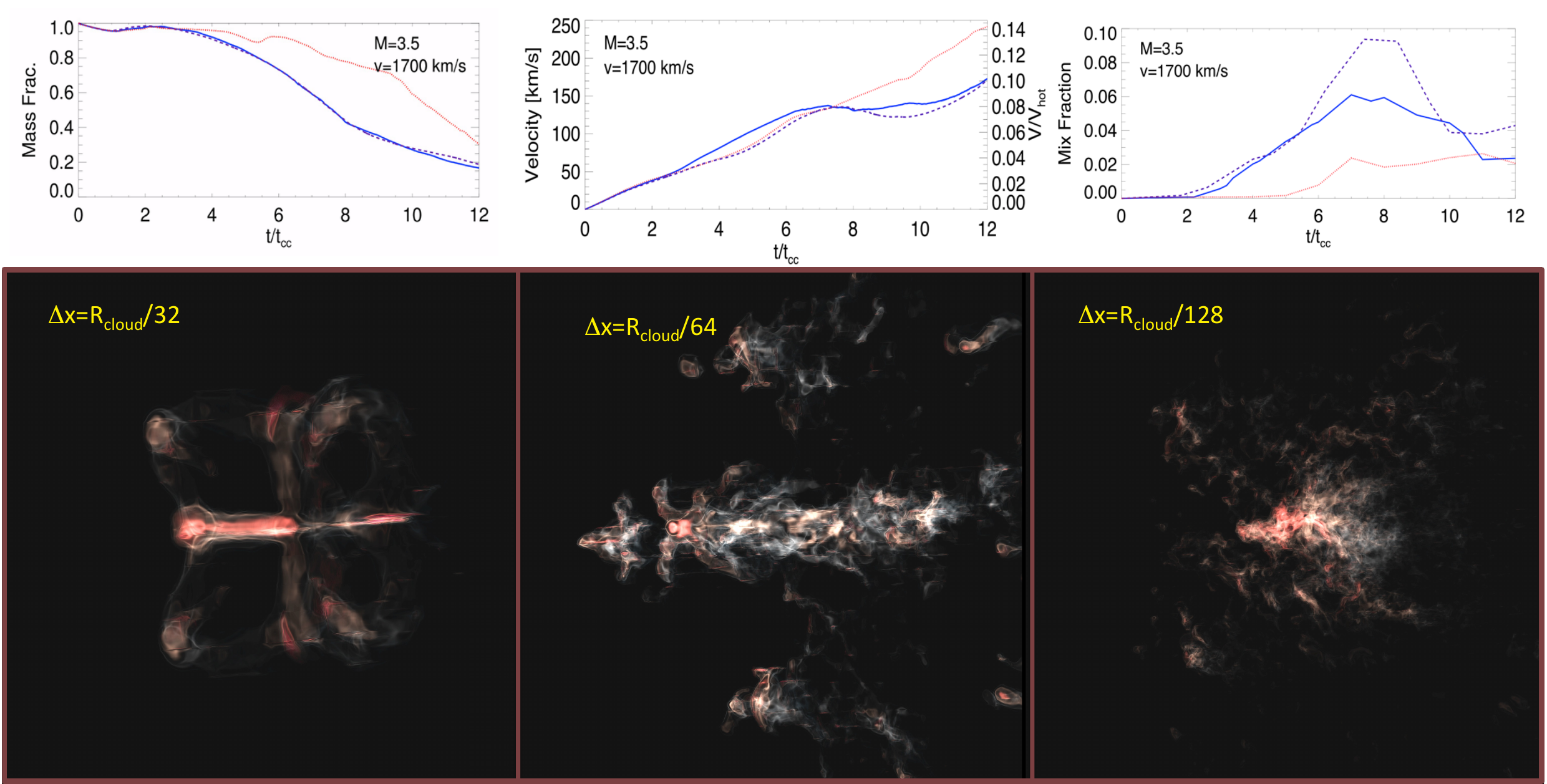}}
\caption{Comparison between $M_{\rm hot}=3.5,$ $v_{\rm hot} = 1700$ km/s, runs with maximum resolutions of $R_{\rm cloud}/32,$   $R_{\rm cloud}/64,$ and $R_{\rm cloud}/128.$ {\em Top Left:} Mass evolution in the fiducial run (blue solid lines) versus the high-resolution run (purple dashed lines) and the low-resolution run (red dotted lines).  {\em Top Center:} Velocity evolution in the three runs, with lines as the top left panel. {\em Top Right:} Evolution of the mixed fraction, $f_{\rm mix},$ in each of the three runs.  {\em Bottom:} Rendered plots of log density from the runs at a time $t_{\rm 75} = 5.8 t_{\rm cc},$ with two surfaces plotted per decade over the range used in the slice plots above: $\log_{10} \rho/({\rm g \, cm}^{-3}) = -25$ to $\log_{10} \rho/({\rm g \, cm}^{-3}) = -22.5.$ The plots are constructed with the camera pointed at the center of the cloud, viewing from a position $(x,y,z)=(-3.3,-0.2,-2.0) R_{\rm cloud}$, and were generated using the yt toolkit (Turk et al. 2011; http://yt-project.org).\\}
\label{fig:Convergence}
\end{figure*}

As a test of resolution effects, we carried out two additional $M_{\rm hot}=3.5$ and $v_{\rm hot} = 1700$ km/s runs: a high-resolution run with a maximum resolution of $\Delta x=R_{\rm cloud}/128,$  and a low-resolution run with a maximum resolution of $\Delta x=R_{\rm cloud}/32,$
as compared to our fiducial $\Delta x=R_{\rm cloud}/64$ resolution.  Note that because we force all our runs to maintain the maximum resolution in the full volume surrounding the cloud, the high-resolution run maintains twice the resolution of the fiducial runs in all important regions of the calculation, while the low-resolution run has half the resolution of the fiducial runs throughout the most important regions.   Our choice to force the regions around the clouds to the maximum resolution also means the three runs vary dramatically in the computer time needed to complete them.  To reach $t= 12 t_{\rm cc},$  the low-resolution run took only $\approx$ 3k CPU hours, the fiducial run took $\approx$ 24k CPU hours, and the high-resolution run took  $\approx$ 260k CPU hours.

 The upper left panel of Figure \ref{fig:Convergence} shows the evolution of $F_{1/3}$, the fraction of total mass at or above 1/3 the original cloud density, in each of these runs.   The agreement between the fiducial and the high-resolution runs is excellent, with $F_{1/3}(t)$ being almost indistinguishable for most of the evolution.    On the other hand, the low-resolution mass evolution is significantly different, with the cloud taking much longer to be disrupted.   This highlights the importance of maintaining adequate resolution when studying the interactions described here.

The evolution of the cloud velocity, shown in the top center panel of Figure \ref{fig:Convergence}, is also very similar between the fiducial and the high-resolution runs, although, in this case, there are minor differences.   While $v_{\rm cloud}(t)$ matches almost exactly between the two runs during the initial stages, at later times the cloud in the higher resolution run reaches slightly lower velocities than the cloud in the fiducial run.  These differences, in turn, appear to be related to the evolution of material stripped from the core of the cloud. Finally, the late time history of the low-resolution run is much different than both of these runs, reaching larger final velocities due to the underestimate of cloud disruption that occur in this simulation.

In order to quantify the evolution of stripped material in each of these runs, we calculated the `mixing-fraction' described in Xu \& Stone (1995) and Orlando \etal (2005), equation (19).  This is defined as 
\be
f_{\rm mix} = \frac{1}{m_{\rm cloud,0}} \int_{V(0.1 < C_{\rm cloud} < 0.9)} dV \,  C_{\rm cloud} \, \rho,
\ee
where $m_{\rm cloud,0}$ is the original mass of the cloud, and the integral is computed over all zones in the simulation in which the mass fraction of the
tracer field $C_{\rm cloud}$ is between 0.1 and 0.9.  In this case, the fiducial and the high-resolution runs track each other well initially, but diverge strongly
after $t \approx 5 t_{\rm cc},$ with the high-resolution run showing a larger mixed fraction than the fiducial case.   The differences between the fiducial and low-resolution run are even stronger, with the low-resolution $f_{\rm mix} \approx 0$ before $5 t_{\rm cc}$ and about half of the fiducial $f_{\rm mix}$ for most of the subsequent evolution. 

The lower panels of Figure \ref{fig:Convergence} show rendered images of the three runs at $5.8 t_{\rm cc},$ the time at which 75\% of the mass is stripped away in the fiducial and high-resolution cases. Here we see that the distribution of the stripped material is significantly different between all three runs. In the fiducial run, the stripped material appears as spray of dense knots that maintain significant density contrasts out to large distances.  On the other hand, the high-esolution run displays a smaller, wispier distribution of stripped material, consistent with better resolved mixing between the two media.    Finally the low-resolution run shows most of the stripped mass collected up in four clumps arranged in an X pattern than is aligned with the diagonals of the simulation grid.   

These variations in morphology help to explain the differences seen in the evolution plots.   The under-resolution of mixing in the $\Delta x = R_{\rm cloud}/32$ run means that the retention of gas by the cloud is overestimated,   leading to high mass fractions and low $f_{\rm mix}$ values throughout the simulation.   Furthermore, because the cloud remains coherent for longer times,  $v_{\rm cloud}(t)$  is the higher in this simulation than in the higher resolution runs.  In the $\Delta x = R_{\rm cloud}/64,$ on the other hand, the cloud evolution is tracked well, but the stripped gas mixing is underestimated.  Thus the more clumpy distribution in the fiducial case includes some fast moving material with densities above 1/3 of the initial cloud density, and this material lasts longer than it does in the high-resolution case, slightly increasing the measured cloud velocity.     Finally, the evolution of $f_{\rm mix}$ in the high-resolution run shows a larger fraction of mixed material that the other runs, as more cells are made up of wispy, $0.1 < C_{\rm cloud} < 0.9$ material.

Computing mixing fractions for the $\Delta x = R_{\rm cloud}/64$ with other Mach numbers and streaming velocities, we find that $f_{\rm mix}$ varies from as little as 1\% to as much as 30\%.  In general, the higher the Mach number and the lower streaming velocity, the higher $f_{\rm mix},$ but these trends are noisy and uncertain.  Thus while the $\Delta x =  R_{\rm cloud}/64$ resolution adopted in our fiducial runs appears to be sufficient to measure the mass loss and velocity evolution in which we are most interested here, even higher resolutions are required to accurately track the  structure of  stripped material, which mixes into the hot medium at significant distances from the central cloud (\eg Marinacci \etal 2010; 2011; Kwak \etal 2011; Henley \etal 2012).

\section{Conclusions}

Galaxy outflows are known to play a key role in the history of galaxy formation, but studies of their properties are limited by the fact that the cold-cloud material that is easiest to observe is also the most difficult to understand.   As a first step in overcoming this limitation, we have conducted a suite of adaptive-mesh refinement simulations that include radiative cooling and track the evolution of such clouds under the full range of conditions relevant for galaxy outflows.  By adopting large simulation volumes, carefully changing frame, and maintaining the highest resolution in the regions in which it is most needed, we have been able to accurately track these clouds for much longer times than previous simulations.  

This is particularly important for highly supersonic flows, in which the disruption of clouds by the KH instability is strongly suppressed. We find that cloud disruption occurs at much later times in this case than in subsonic interactions, but still it occurs much sooner than expected from laboratory experiments of supersonic shear layers. Rather than finding lifetimes $\propto t_{\rm cc} M^{-1}$ as expected from eq.\ (\ref{eq:Slessor}), we find that the times at which that cloud loses 90\% of its mass ($t_{90}$), 75\% of its mass ($t_{75}$),  50\% of its mass ($t_{50}$),  $t_{50} $ and 25\% of its mass ($t_{25}$) are given by 
\be
t= \alpha t_{\rm cc} \sqrt{1+M_{\rm hot}},
\label{eq:tfit2}
\ee
where $\alpha $= 1.75, 2.5, 4, and 6, at $t_{90}$ $t_{75}$ $t_{50} $ and $t_{25}$, respectively. 

The reason for this scaling lies in the structure is the structure of the Mach cone that forms around the cloud.  While the normal shock at the head of the cloud increases the pressure there by  a factor $\approx 1+M^2,$ the downstream pressure rises only by a factor $1+M_{\rm hot}$ is it experiences only an oblique shock.   Thus the cloud becomes compressed in the direction normal to the flow,  but the pressure gradient in the streamwise direction causes it to expand in this direction by a factor  $\propto t^2/[t^2_{\rm cc} (1+M_{\rm hot})].$  This in turn leads to a ``cometary'' appearance at intermediate times and finally disrupts the cloud completely on a timescale  $\propto t_{\rm cc} \sqrt{1+M_{\rm hot}}.$ 

We also find that the cloud velocity is primarily set by the momentum imparted by the impinging flow, 
as described by eq.\ (\ref{eq:vcoft}).
Our simulations show that deviations from this estimate depend almost solely on the Mach number. We find a fit for the cloud velocity as a function of the exterior Mach number given by 
\be
v_{\rm cloud}(t) = \frac{3 v_{\rm hot} \left( \alpha \sqrt{1+M_{\rm hot}} \right)}{4 \chi_0^{1/2}} \frac{1.8}{(1+M_{\rm hot})^\beta}, 
\ee
where again $\alpha$ = 1.75, 2.5, 4, and 6, at $t_{90},$ $t_{75},$ $t_{50},$ and $t_{25},$ respectively, and where $\beta = 0.8$ fits best at $t_{90}$ and $t_{75}$ and $\beta=1$ fits best at later times. These scalings can be understood in terms of changes in the effective radius of the cloud perpendicular to the direction of the flow.  The $\beta = 0.8$ fit at early times represents cloud compressions that are somewhat biased to the direction perpendicular to the  streamwise direction, while the the $\beta=1.0$ fits at late time represents compressions that are almost completely perpendicular to the streamwise direction in which the cloud is stretched.

Integrating our fit to $v_{\rm cloud}$ out to the measured disruption times, we find $d_{\rm cloud} = .65 \alpha^2 R_{\rm cloud} (1+M)^{(1-\beta)},$ with $\alpha$ and $\beta$ defined as above.  Because $\beta \approx 1,$  the distance the cloud travels as a function of remaining mass fraction is roughly constant over a large range of external conditions, with the clouds becoming completely disrupted as they move beyond $d_{\rm cloud} \approx  40 R_{\rm cloud.}$ These distances are large enough that, given initial cloud radii $\lesssim $ 100 parsecs, cold clouds can travel from near the edge of the driving region to distances of a few kpc, to be observed as absorption features superimposed on the spectrum of there host galaxies. On the other hand, unless conduction and magnetic effects are able to preserve clouds for much longer than seen our current simulations, $\approx 10^4$K gas observed at $\approx 100$ kpc distances around galaxies can not be directly associated with $10^4$K ejected material.

In order to test resolution effects, we carried out runs with $M_{\rm hot}=3.5$ and $v_{\rm hot} = 1700$ km/s, and maximum resolution levels of $\Delta x  = R_{\rm cloud}/32$ and  $\Delta x  =  R_{\rm cloud}/128,$  as opposed to the fiducial values of  $\Delta x  =  R_{\rm cloud}/64$.   We find that the $\Delta x =R_{\rm cloud}/32$ results vary strongly from those of the other two runs, indicating that $\Delta x  = R_{\rm cloud}/32$ is not adequate to obtain reliable results. On the other hand, the cloud mass and velocity evolution do not change significantly between the $\Delta x = R_{\rm cloud}/64,$ and $\Delta x= R_{\rm cloud}/128$ runs, although the mass fraction of the mixing layers between the fluids increases and their structure becomes more complex at very high resolution.

The mass and velocity scalings found above can be implemented in numerical simulations of galactic outflows on larger scales. As the evolution of cold clouds in such simulations cannot be adequately modeled directly, subgrid models that use our scalings for mass loss and cloud velocity could represent a promising path forward to to model the thermal evolution of the multi-phase material in outflows.  However, before this can be carried out with confidence, we must first understand how thermal conduction and magnetic fields impact the cloud and change its evolution relative to the hydrodynamic case with cooling. These physical effects will be simulated in forthcoming work.

\acknowledgements

We would like to thank Romeel Dav\' e, Paul E. Dimotakis,  Timothy Heckman, Crystal Martin, Eve Ostriker, and Todd Thompson for helpful discussions. ES gratefully acknowledge the Simons Foundation for funding the workshop {\em Galactic Winds: Beyond Phenomenology} which helped to inspire this work. He also gratefully acknowledges Joanne Cohn, Eliott Quatert, and the UC Berkeley Theoretical Astronomy Center,  Uros Seljak and the Lawrence Berkeley National Lab Cosmology group, and the organizers of the {\em Gravity's Loyal Opposition: The Physics of Star Formation Feedback} at the Kavli Institute for Theoretical Physics, for hosting him during the period when much of this work was carried out.  We would also like to acknowledge the Texas Advanced Computing Center (TACC) at The University of  Texas at Austin, and  the Extreme Science and Engineering Discovery  Environment (XSEDE) for providing HPC resources via grant TG-AST140004 that have contributed to the research results reported within this paper.  Some simulations were also run on the JUROPA supercomputer at the Juelich Centre for Supercomputing under project HHB03. The FLASH code was developed in part by the DOE-supported  Alliances Center for Astrophysical Thermonuclear Flashes (ASC) at the University of Chicago. ES was supported by NSF grants AST11-03608 and PHY11-25915, and NASA theory grant NNX09AD106.

\appendix

\section{Shock Jump Conditions}

\begin{figure}
\centering
\includegraphics[width=1.0\columnwidth]{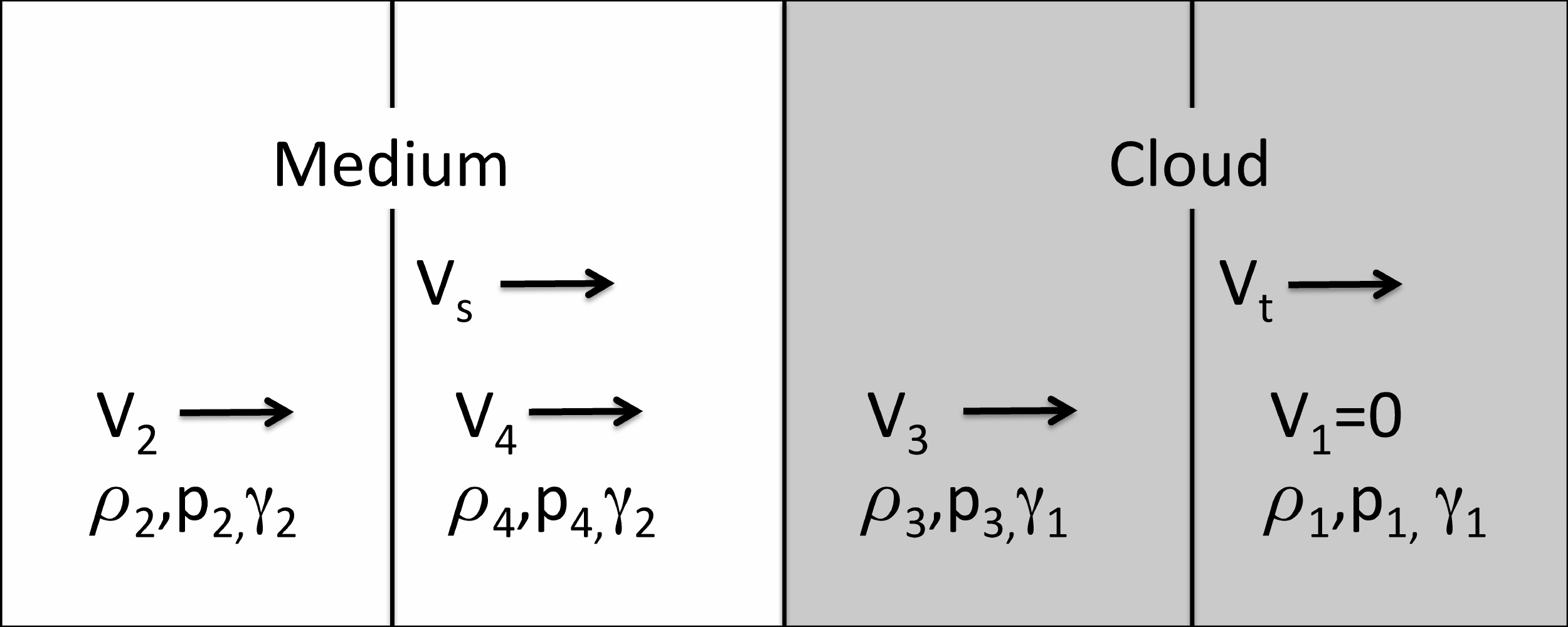}
\caption{Definition of coordinates for the shock.}
\label{fig:align2}
\end{figure}

The are four distinct regions that describe the interaction when the shock first hits the front of the cloud: the undisturbed cloud (region 1), the unshocked exterior medium  (region 2), the region behind the shock that is transmitted through the cloud  (region 3), and the region behind the shock that is reflected back into the exterior medium  (region 4).  These are illustrated in Figure \ref{fig:align2} (see also Silk \&Solinger 1974; Hester \etal 1994).
Within the cloud, in the frame in which the cloud is initially stationary, we have:
\be
\frac{\rho_1}{\rho_3} = \frac{v_t-v_3}{v_t} = \frac{M_t- M'_3}{M_t} = \frac{(\gamma_1-1) M_t^2 +2} {(\gamma_1 + 1) M_t^2},
\label{eq:transmit}
\ee
and
\be
\frac{p_3}{p_1} = \frac{2 \gamma_1 M_t^2}{\gamma_1 +1} - \frac{\gamma_1 -1}{\gamma_1+1},
\label{eq:p31}
\ee
where $v_t$ is the velocity of the transmitted shock, $M_t \equiv v_t/c_{s,1}$, and $M_3' \equiv v_3/c_{s,1}$, and where we use $'$ to denote the ratio of the velocity behind the transmitted shock with the sound speed in front of the shock. 
In this same frame, within the exterior medium we have
\be
\frac{\rho_2}{\rho_4} = \frac{v_4-v_s}{v_2-v_s} =  \frac{M_4'-M_s}{M_2 - M_s} = \frac{(\gamma_2-1) (M_2-M_s)^2 +2}{(\gamma_2 + 1) (M_2-M_s)^2},
\label{eq:incomming}
\ee
and
\be
\frac{p_4}{p_2} = \frac{2 \gamma_2 (M_2-M_s)^2}{\gamma_2 +1} - \frac{\gamma_2-1}{\gamma_2+1},
\label{eq:p42}
\ee
where $M_s= v_s/c_{s,2}$, $M_2 = v_2/c_{s,2},$ and $M_4'=v_4/c_{s,2},$ and again the $'$ denotes the ratio of the post shock velocity (in region 4) with the  pre-shock sound speed (in region 2). 
We also have that $p_3=p_4$, $v_3=v_4,$
and, because the pre-shock exterior material the cloud are assumed to be initial pressure equilibrium, $p_3/p_1 = p_4/p_2$.   

These relations allow us to  equate eq.\ (\ref{eq:p31}) and eq.\ (\ref{eq:p42}) to relate the incoming flow to shock passing through the cloud for arbitrary  choices of $\gamma_1$ and $\gamma_2$.  
If $\gamma_1=\gamma_2=\gamma,$ this becomes particularly simple.  In this case
we have that $M_2-M_s = M_t $ and so from eqs.\ (\ref{eq:transmit}) and (\ref{eq:incomming}), $\rho_1/\rho_3=\rho_2/\rho_4,$
such that $\rho_3/\rho_4 = \chi_0.$
This means
\be
M_t- M'_3 = M_4'-M_s  =  M'_4-M_2+M_t,  
\ee
and $M_2 = M_3'(1+ \chi_0^{-1/2}).$
We then rearrange eq.\ (\ref{eq:transmit}) by multiplying both sides by $(\gamma+1)M_t^2$ and plugging in for $M_3'$ to obtain
\be
M_t^2-1 =  \tilde M_2  M_t, 
\ee
where $\tilde M_2 \equiv  M_2 \frac{\gamma+1}{2 (1+\chi_0^{-1/2})}.$
Solving for $M_t,$ this gives
\be
M_t = \tilde M_2 \frac{ 1 + ( 1 + 4 \tilde M_2^{-2})^{1/2}}{2},
\label{eq:Mt}
\ee
such that $v_t \approx v_2/\chi_0^{1/2}$ in the high Mach number limit.   Combined with eq.\ (\ref{eq:p31}) this gives the post-shock pressure in the cloud in terms of the original pressure. Technically, these relations apply only when the Mach number of the incoming flow is 1 or above, but in practice extending this to lower velocities gives reasonable estimates.

\bibliographystyle{apjsingle}

\end{document}